\documentclass[12pt,preprint]{article}
\usepackage{amssymb}
\usepackage{amsmath}
\usepackage{graphics}
\usepackage{epsfig}

\newcommand{\eqb}{\begin{equation}}
\newcommand{\eqe}{\end{equation}}
\newcommand{\dmb}{\begin{displaymath}}
\newcommand{\dme}{\end{displaymath}}

\newcommand{\eab}{\begin{eqnarray}}
\newcommand{\eae}{\end{eqnarray}}

\newcommand{\be}{\begin{equation}}
\newcommand{\ee}{\end{equation}}

\begin{document}

\begin{titlepage}
\begin{flushright} 
\end{flushright}
\vspace{0.6cm}

\begin{center}
\Large{Mixing of scalar tetraquark and quarkonia states
in a chiral approach}
\vspace{1.5cm}

\large{Francesco Giacosa}

\end{center}
\vspace{1.5cm} 

\begin{center}
{\em Institut f\"ur Theoretische Physik\\ 
Universit\"at Frankfurt\\ 
Johann Wolfgang Goethe - Universit\"at\\ 
Max von Laue--Str. 1\\ 
60438 Frankfurt, Germany\\
e-mail: giacosa@th.physik.uni-frankfurt.de}
\end{center}
\vspace{1.5cm}

\begin{abstract}
A chiral  invariant Lagrangian
describing the tetraquark-quarkonia interaction is considered at the leading 
and subleading order in the large-$N_{c}$  expansion. Spontaneous
chiral symmetry breaking generates mixing of scalar tetraquark and quarkonia states
and non-vanishing tetraquark condensates. In particular, the mixing strength 
is related to the decay strengths of tetraquark states into pseudoscalar mesons. The results show that 
scalar states below 1 GeV are mainly
four-quark states and the scalars between 1 and 2 GeV quark-antiquark states, probably mixed with 
the scalar glueball in the isoscalar sector.
\end{abstract} 

\end{titlepage}

\bigskip

\bigskip

\section{Introduction}

The spectroscopic interpretation of the scalar states below 1 GeV represents
an important issue of modern hadronic physics. It is not yet clear if the
dominant contribution to their wave function constitutes of quarkonia,
mesonic molecules or Jaffe's tetraquark states. In turn, this subject is
strongly connected to the nature of the scalar states above 1 GeV (we refer
to the review papers \cite{amslerrev,closerev,exotica}).

Various interpretations have been proposed in the literature about the
scalar resonances below and above 1 GeV. According to the most popular
scenario, one interprets the isovector and isotriplet resonances $%
a_{0}(1450) $ and $K(1430)$ as the ground-state quark-antiquark bound
states. The three isoscalar resonances $f_{0}(1370)$, $f_{0}(1500)$ and $%
f_{0}(1710)$ are an admixture of two isoscalar quarkonia and bare glueball
configurations (we refer to \cite%
{amslerrev,closerev,close95,weing,gutsche,closekirk,giacosa,giacosachiraluno,giacosachiral,cheng}
and Refs. therein; recently the inclusion of hybrids in the mixing scheme
has been performed in Ref. \cite{heli}). As a consequence, the scalar states
below 1 GeV ($f_{0}(600),$ $k(800),$ $f_{0}(980)$ and $a_{0}(980)$) must be
something else, like (loosely bound) mesonic molecular states \cite%
{mesonicmol1,mesonicmol2}, dynamical generated resonances \cite{oller} or
Jaffe's tetraquark states \cite{amslerrev,exotica,jaffe,maiani,tq}. It is
indeed possible that an interplay of these three possibilities takes place.

The tetraquark states, whose building blocks are a diquark ($q^{2}$) and an
antidiquark ($\overline{q}^{2}$), play a central role in this paper.
Calculations based on one-gluon exchange \cite{amslerrev,onegluon},
instantons \cite{inst1,inst2}, Nambu Jona--Lasinio model (NJL) \cite{weise}
and Dyson-Schwinger equation (DSE) \cite{maris} support a strong attraction
among two quarks in a color antitriplet ($\overline{3}_{C}$), a flavor
antitriplet ($\overline{3}_{F}$) and spinless configuration \cite%
{amslerrev,jaffe} (color and flavor triplets are realized for an
antidiquark). Naively speaking, such a scalar diquark `behaves like an
antiquark' from a flavor (and color) point of view, thus a nonet of light
scalar tetraquark states naturally emerges in this context. Support for the
existence of Jaffe's states below 1 GeV is in agreement with the Lattice
studies of Refs. \cite{jaffelatt,okiharu,mathur}.

In the recent work of Ref. \cite{tq} the present author analyzed the strong
and the electromagnetic decays of the light scalar states $\{f_{0}(600),$ $%
k(800),$ $f_{0}(980)$ and $a_{0}(980)\}$\footnote{%
The resonance $k(800)$ is now listed in the compilation of the Particle Data
Group \cite{pdg} but it still needs confirmation and is omitted from the
summary table. The resonance is also found in many recent theoretical and
experimental works (\cite{oller,vanbeveren,ishida,black,bugg,buggexp06} and
Refs. therein).} interpreted as Jaffe's tetraquark states, which naturally
account for the mass degeneracy of $f_{0}(980)$ and $a_{0}(980)$ and their
large $\overline{K}K$ decay strength. The dominant (Fig. 1.a) and the
subdominant (Fig. 1.b) decay mechanisms in the large-$N_{c}$ expansion,
respectively proportional to the decay strengths $c_{1}$ and $c_{2}$, have
been systematically taken into account in an effective $SU_{V}(3)$-invariant
interaction Lagrangian.

In the present work we extend the model of Ref. \cite{tq}, which was built
under the requirement of flavor symmetry $SU_{V}(3)$, by considering
invariance under the chiral group $SU_{R}(3)\times SU_{L}(3)$. The explicit
inclusion of a scalar quarkonia nonet, lying between 1 and 2 GeV (see
discussion in Section 2.1) as the chiral partner of the pseudoscalar nonet,
and the inclusion of the pseudoscalar diquark, as the chiral partner of the
scalar diquark, are required. As in \cite{tq} we keep the leading and the
subleading terms in the large-$N_{c}$ expansion.

As a consequence of chiral symmetry breaking, mixing among tetraquark and
quarkonia states takes place. The most important theoretical result of the
present work is the possibility to relate the mixing strength between the
scalar tetraquark and quarkonia nonets to the tetraquark decay strengths $%
c_{1}$ and $c_{2}$ of Fig. 1 and to the pion and kaon decay constants.
Furthermore, the tetraquark-quarkonia mixing in the scalar sector is
responsible for the emergence of non-vanishing tetraquark condensates.

The connection of the decay strengths $c_{1}$ and $c_{2}$ to the mixing
allows us to evaluate its strength. As a result we find that the tetraquark
assignment for the light scalar states is consistent: by analyzing the
isovector channel the resonance $a_{0}(980)$ has a dominant tetraquark
content; the quarkonium amount in its spectroscopic wave function turns out
to be relatively small ($\lesssim 10\%$). An analogous result is obtained in
the kaonic sector.

The use of chiral Lagrangian for the analysis of tetraquark-quarkonia mixing
has been studied in Refs. \cite{shechter,fariborz,napsuciale,fariborzlast},
where sizable admixtures in the scalar physical resonances below and above 1
GeV are found. In the present work a different chiral Lagrangian is utilized
and only the scalar (and not the pseudoscalar) diquarks are considered as
basic constituent for low-energy mesonic resonances. Our results point to a
smaller mixing strength and thus to a substantial separation of four-quark
states below 1 GeV and quarkonia states above 1 GeV.

The paper is organized as follows. In Section 2 the model is constructed: we
recall the basics of the chiral treatment of the scalar and pseudoscalar
nonets, we introduce the scalar diquark and briefly review Ref. \cite{tq},
we describe the pseudoscalar diquark and write down the chiral invariant
tetraquark-quarkonia interaction Lagrangian. In section 3 the
phenomenological implications are studied: the scalar tetraquark-quarkonia
mixing and the magnitude of the tetraquark condensates. In section 4 we
present the summary and the conclusions.

\section{Set-up of the model}

\subsection{Quarkonia nonets}

We briefly recall the basic elements for the set up of the pseudoscalar and
the scalar quarkonia nonets. At a microscopic level one has the quark field $%
q_{i}(x)$ with $i=u,d,s.$ The right and left spinors are given by:%
\begin{eqnarray}
q_{i,R} &=&P_{R}q_{i},\text{ }q_{i,R}^{\dag }=q_{i}P_{R},\text{ }\overline{q}%
_{i,R}=\overline{q}_{i}P_{L} \\
\text{ }q_{i,L} &=&P_{L}q_{i},\text{ }q_{i,L}^{\dag }=q_{i}P_{L},\text{ }%
\overline{q}_{i,L}=\overline{q}_{i}P_{R},
\end{eqnarray}%
where $P_{R}=\frac{1}{2}\left( 1+\gamma _{5}\right) $ and $P_{L}=\frac{1}{2}%
\left( 1-\gamma _{5}\right) .$

The $SU_{R}(3)\times $ $SU_{L}(3)$ transformation on the quark fields is
defined as:%
\begin{equation}
q_{i}=q_{i,R}+q_{i,L}\rightarrow R_{ij}q_{j,R}+L_{ij}q_{j,L}\text{ with: }%
R\in SU_{R}(3)\text{, }L\in SU_{L}(3).  \label{chiralt}
\end{equation}%
Out of quark fields one can build up operators (currents) with the correct
quantum numbers of the physical resonances. In fact, at a composite level
one deals with mesons, which have the same transformation properties of the
underlying quark currents. In Table 1 we summarize the properties of the
pseudoscalar and scalar quarkonia Hermitian matrices $\mathcal{P}$ and $%
\mathcal{S}$ and of the matrix $\Sigma =\mathcal{S}+i\mathcal{P}$: the
corresponding matrix elements and the components in the Gell-Mann basis
(denoted as `currents' in Table 1), the transformations under parity (P),
charge conjugation (C), $SU_{V}(3)$ (occurring for $R=L=U$ in Eq. (\ref%
{chiralt})), chiral $SU_{R}(3)\times $ $SU_{L}(3)$ and $U_{A}(1)$ (occurring
for $q_{i}\rightarrow e^{i\nu \gamma _{5}}q_{i},$ i.e. $q_{i,R}\rightarrow
e^{i\nu }q_{i,R}$ and $q_{i,L}\rightarrow e^{-i\nu }q_{i,L}$) are reported.

\begin{center}
\textbf{Table 1}: Summary of the properties of $\mathcal{P},$ $\mathcal{S}$
and $\Sigma $.

\begin{tabular}{|c|c|c|c|}
\hline
& $\mathcal{P}=\frac{1}{\sqrt{2}}\sum_{i=0}^{8}P^{i}\lambda _{i}$ & $%
\mathcal{S=}\frac{1}{\sqrt{2}}\sum_{i=0}^{8}S^{i}\lambda _{i}$ & $\Sigma =%
\mathcal{S}+i\mathcal{P}$ \\ \hline
Matrix Elements & $\mathcal{P}_{ij}\equiv \overline{q}_{j}i\gamma ^{5}q_{i}$
& $\mathcal{S}_{ij}\equiv \overline{q}_{j}q_{i}$ & $\Sigma _{ij}\equiv 2%
\overline{q}_{j,R}q_{i,L}$ \\ \hline
Currents & $P^{i}\equiv \overline{q}i\gamma ^{5}\frac{\lambda _{i}}{\sqrt{2}}%
q$ & $S^{i}\equiv \overline{q}\frac{\lambda _{i}}{\sqrt{2}}q$ & $\Sigma
^{i}\equiv 2\overline{q}_{R}\frac{\lambda _{i}}{\sqrt{2}}q_{L}$ \\ \hline
P & $-\mathcal{P(}x^{0},-\mathbf{x}\mathcal{)}$ & $\mathcal{S(}x^{0},-%
\mathbf{x}\mathcal{)}$ & $\Sigma ^{\dagger }\mathcal{(}x^{0},-\mathbf{x}%
\mathcal{)}$ \\ \hline
C & $\mathcal{P}^{t}$ & $\mathcal{S}^{t}$ & $\Sigma ^{t}$ \\ \hline
$SU_{V}(3)$ & $U\mathcal{P}U^{\dagger }$ & $U\mathcal{S}U^{\dagger }$ & $%
U\Sigma U^{\dagger }$ \\ \hline
$SU_{R}(3)\times $ $SU_{L}(3)$ & $\frac{1}{2i}\left( L\Sigma R^{\dagger
}-R\Sigma ^{\dagger }L^{\dagger }\right) $ & $\frac{1}{2}\left( L\Sigma
R^{\dagger }+R\Sigma ^{\dagger }L^{\dagger }\right) $ & $L\Sigma R^{\dagger
} $ \\ \hline
$U_{A}(1)$ ($q_{i}\rightarrow e^{i\nu \gamma _{5}}q_{i}$) & $\frac{1}{2i}%
\left( e^{-2i\nu }\Sigma -e^{2i\nu }\Sigma ^{\dagger }\right) $ & $\frac{1}{2%
}\left( e^{-2i\nu }\Sigma +e^{2i\nu }\Sigma ^{\dagger }\right) $ & $%
e^{-2i\nu }\Sigma $ \\ \hline
\end{tabular}
\end{center}

\bigskip

Following \cite{black} and Refs. therein, which we refer to for a careful
treatment, we introduce the Lagrangian $\mathcal{L}_{\Sigma }$ 
\begin{equation}
\mathcal{L}_{\Sigma }=\frac{1}{4}Tr\left[ \partial _{\mu }\Sigma \partial
^{\mu }\Sigma ^{\dag }\right] -V_{0}(\Sigma ,\Sigma ^{\dag })-V_{SB}(\Sigma
,\Sigma ^{\dag })  \label{lsigma}
\end{equation}%
($Tr$ denotes trace over flavor) which describes the dynamics of the
pseudoscalar and scalar quarkonia mesons. As usual, $V_{0}$ represents the
chiral invariant potential while $V_{SB}$ encodes symmetry breaking due to
the non-zero current quark masses\footnote{%
Notice that we are considering nonets of states and not only octects (the
sum in Table 1 runs from $i=0,...,8$). Thus, in $V_{SB}$ also $U_{A}(1)$
breaking, mixing and large-$N_{c}$ suppressed terms are (implicitly)
included.}. In the present work we are not concerned with the detailed
description of the properties of the potentials $V_{0}$ and $V_{SB}$. What
is important for us is spontaneous chiral symmetry breaking ($\chi SB$),
that is the minimum of the potential $V_{0}+V_{SB}$ is realized for non-zero
vacuum expectation values ($vev$'s):%
\begin{equation}
\left\langle \mathcal{S}_{ij}\right\rangle =\alpha _{i}\delta _{ij}\text{
(with }\alpha _{u}=\alpha _{d}\text{: isospin symmetry).}  \label{csb1}
\end{equation}%
The expectation values $\alpha _{i}$ are related to the pion and the kaon
decay constants in a model independent way \cite{black,geffen}:%
\begin{equation}
\left\langle \mathcal{S}_{11}\right\rangle +\left\langle \mathcal{S}%
_{22}\right\rangle =2\alpha _{u}=\sqrt{2}F_{\pi }\text{, }\left\langle 
\mathcal{S}_{11}\right\rangle +\left\langle \mathcal{S}_{33}\right\rangle
=\alpha _{u}+\alpha _{s}=\sqrt{2}F_{K}\text{.}  \label{alfaeq}
\end{equation}%
We use $F_{\pi }=0.0924$ GeV and $F_{K}=1.22F_{\pi }$. This leads us to
shift the matrix $\Sigma $ as:%
\begin{equation}
\Sigma =\mathcal{S}+i\mathcal{P\rightarrow }\Sigma =\Sigma _{0}+\mathcal{S}+i%
\mathcal{P}\text{ where: }\Sigma _{0}=diag\{\alpha _{u},\alpha _{u},\alpha
_{s}\}.  \label{sigma0}
\end{equation}

The pseudoscalar nonet is well established: $\mathcal{P\equiv }\{\pi ,K,\eta
,\eta ^{\prime }\}.$ The identification of the scalar states is
controversial. Some models \cite{weise,scadron,kleefeld,juergen} identify
the resonance $f_{0}(600)$ as the chiral partner of the pion, hence a
quarkonium $\overline{n}n=\sqrt{1/2}(\overline{u}u+\overline{d}d)$. This
assignment encounters a series of well-known problems: (i) in this scheme
the resonances $a_{0}^{0}(980)$ and $f_{0}(980)$ would be respectively $%
\sqrt{1/2}(\overline{u}u-\overline{d}d)$ and $\overline{s}s.$ Their mass
degeneracy is then hard to be explained from their quark content (see also
the different point of view in Refs. \cite{kleefeld,juergen}) (ii) The
strong coupling of $a_{0}(980)$ to $\overline{K}K$ cannot be explained
within this assignment (the points (i)-(ii) are naturally explained when
interpreting the light scalar resonance as mainly Jaffe's four-quark states,
see Refs. \cite{jaffe,exotica,maiani,tq} and next subsection). (iii) The
scalar quarkonia states are p-wave $(L=S=1),$ therefore expected to have a
mass comparable to the p-wave nonets of tensor and axial-vector mesons which
lie well above 1 GeV. (iv) The Lattice results of Refs. \cite%
{mathur,prelovsek} predict a mass for the quarkonium state $u\overline{d}$
about $1.4$-$1.5$ GeV, thus well above 1 GeV (see also the different result
of Ref. \cite{mcneile}). (v) As shown in Ref. \cite{pelaez} (and recently
confirmed in Ref. \cite{pelaezlast} at two-loop order in unitarized Chiral
Perturbation Theory for the resonance $f_{0}(600)$) the large-$N_{c}$
behavior of the masses of the scalar states below 1 GeV is $not$ compatible
with a dominant quarkonium content, thus further pointing to a heavier bare
mass of the latter.

Thus, we expect that the bare quarkonia masses lie above 1 GeV. We will then
analyze the mixing of the quarkonia states with the (lighter) four-quark
states in Section 3.

\subsection{Scalar diquark and corresponding tetraquark states}

We turn our attention to the scalar diquark current. To this end we consider
the following scalar flavour-antisymmetric ($\overline{3}_{F}$)
diquark-matrix $D$:%
\begin{equation}
D_{ij}\equiv \sqrt{\frac{1}{2}}\left( q_{j}^{t}C\gamma
^{5}q_{i}-q_{i}^{t}C\gamma ^{5}q_{j}\right) =\sum_{i=1}^{3}\varphi _{i}A^{i}
\label{d}
\end{equation}%
\begin{equation}
\left( A^{i}\right) _{jk}=\varepsilon _{ijk},\text{ }\varphi _{i}=\sqrt{%
\frac{1}{2}}\varepsilon _{ijk}q_{j}^{t}C\gamma ^{5}q_{k}\text{ ,}
\label{phii}
\end{equation}%
where the superscript $t$ refers to transposition in the Dirac space. Color
indices, formally identical to the flavor ones ($\overline{3}_{C}$), are
understood. We refer to the quantities $\varphi _{i},$ arising from the
decomposition of $D$ in the basis of the antisymmetric matrices $A^{i},$ as
the scalar diquark currents and to the Hermitian conjugate $\varphi
_{i}^{\dagger }$ as the scalar antidiquark currents.

In terms of flavour the currents $\varphi _{i}$ read:%
\begin{equation}
\varphi _{1}=\sqrt{\frac{1}{2}}[d,s]\leftrightarrow \overline{u},\text{ }%
\varphi _{2}=-\sqrt{\frac{1}{2}}[u,s]\leftrightarrow \overline{d},\text{ }%
\varphi _{3}=\sqrt{\frac{1}{2}}[u,d]\leftrightarrow \overline{s}
\label{corr}
\end{equation}%
where the correspondence $\leftrightarrow $ refers to the fact that a
diquark in the flavor (and color) antisymmetric decomposition behaves like
an antiquark, as already anticipated in the Introduction.

The spinor structure of the kind $q^{t}C\gamma ^{5}q$ corresponds to a
diquark with parity $+1$ (ergo to $L=S=0$). Schematically:

\begin{equation}
\left\vert qq\right\rangle _{L=S=0}=\left\vert \text{space: }%
L=0\right\rangle \left\vert \text{spin: }S=0\right\rangle \left\vert \text{%
color: }\overline{3}_{C}\right\rangle \left\vert \text{flavor: }\overline{3}%
_{F}\right\rangle ;\text{ }J^{P}=0^{+}.  \label{diqpiu}
\end{equation}%
As discussed in the Introduction the scalar diquark $\left\vert
qq\right\rangle _{L=S=0}$ forms a compact and stable object, as one-gluon
exchange, instanton-based calculations, NJL and DSE approaches show,
rendering it a good constituent for light meson (and baryon) spectroscopy 
\cite{exotica}.

In Table 2 we recall the microscopic decomposition of the elements of the
diquark matrix $D$ and the corresponding currents $\varphi _{i}$ of Eq. (\ref%
{d}) and the properties under $SU_{V}(3),$ parity and charge-conjugation
transformations.

\begin{center}
\textbf{Table 2: }Properties of the scalar diquark matrix $D$ and components 
$\varphi _{i}$.

$%
\begin{tabular}{|c|c|c|}
\hline
& $D=\sum_{i=1}^{3}\varphi _{i}A^{i}$ & $\varphi _{i}$ \\ \hline
Elements/Currents & $D_{ij}=\sqrt{\frac{1}{2}}\left( q_{j}^{t}C\gamma
^{5}q_{i}-q_{i}^{t}C\gamma ^{5}q_{j}\right) $ & $\text{ }\varphi _{i}=\sqrt{%
\frac{1}{2}}\varepsilon _{ijk}q_{j}^{t}C\gamma ^{5}q_{k}$ \\ \hline
P & $D$ & $\varphi _{i}$ \\ \hline
C & $D^{\dagger }$ & $\varphi _{i}^{\dagger }$ \\ \hline
$SU_{V}(3)$ & $UDU^{t}$ & $\varphi _{k}U_{ki}^{\dagger }$ \\ \hline
\end{tabular}%
$
\end{center}

As one can notice, the $SU_{V}(3)$-transformation of the diquark currents $%
\varphi _{i}\rightarrow \varphi _{k}U_{ki}^{\dagger }$ is exactly analogous
to the $SU_{V}(3)$-transformation of an antiquark: $\overline{q}%
_{i}\rightarrow \overline{q}_{k}U_{ki}^{\dagger }.$ This is the formal way
to express the correspondences in Eq. (\ref{corr}).

The scalar tetraquark nonet is given by the composition of a scalar diquark
and a scalar antidiquark, resulting in the following diquark-current:

\begin{equation}
\mathcal{S}_{ij}^{[4q]}=\varphi _{i}^{\dagger }\varphi _{j},  \label{s4q}
\end{equation}%
where the superscript $[4q]$ refers to four-quark states and avoid confusion
with the scalar quarkonia nonet introduced previously.

In flavor components $\mathcal{S}^{[4q]}$ explicitly reads (from Eqs. (\ref%
{d}) and (\ref{phii})):%
\begin{eqnarray}
\mathcal{S}^{[4q]} &=&\frac{1}{2}\left( 
\begin{array}{ccc}
\lbrack \overline{d},\overline{s}][d,s] & -[\overline{d},\overline{s}][u,s]
& [\overline{d},\overline{s}][u,d] \\ 
-[\overline{u},\overline{s}][d,s] & [\overline{u},\overline{s}][u,s] & -[%
\overline{u},\overline{s}][u,d] \\ 
\lbrack \overline{u},\overline{d}][d,s] & -[\overline{u},\overline{d}][u,s]
& [\overline{u},\overline{d}][u,d]%
\end{array}%
\right)  \label{tet1} \\
&=&\left( 
\begin{array}{ccc}
\sqrt{\frac{1}{2}}(f_{B}[4q]-a_{0}^{0}[4q]) & -a_{0}^{+}[4q] & k^{+}[4q] \\ 
-a_{0}^{-}[4q] & \sqrt{\frac{1}{2}}(f_{B}[4q]+a_{0}^{0}[4q]) & -k^{0}[4q] \\ 
k^{-}[4q] & -\overline{k}^{0}[4q] & \sigma _{B}[4q]%
\end{array}%
\right)  \label{tet2}
\end{eqnarray}%
where in Eq. (\ref{tet2}) we explicitly introduced the tetraquark fields. In
particular, the states $\sigma _{B}[4q]=\frac{1}{2}[u,d][\overline{u},%
\overline{d}]$ and $f_{B}[4q]=\frac{1}{2\sqrt{2}}([u,s][\overline{u},%
\overline{s}]+[d,s][\overline{d},\overline{s}])$ refer to $bare$ (unmixed)
tetraquark scalar-isoscalar states.

In Ref. \cite{tq} the $SU_{V}(3),$ C and P invariant interaction Lagrangian
describing the decay of a tetraquark meson into two pseudoscalar quarkonia
mesons has been introduces as:%
\begin{eqnarray}
\mathcal{L}_{\mathcal{S}^{[4q]}PP} &=&-c_{1}Tr\left[ D\mathcal{P}^{t}D^{\dag
}\mathcal{P}\right] +c_{2}Tr\left[ DD^{\dag }\mathcal{P}^{2}\right]
\label{ltpp} \\
&=&c_{1}\mathcal{S}_{ij}^{[4q]}Tr\left[ A^{j}\mathcal{P}^{t}A^{i}\mathcal{P}%
\right] -c_{2}\mathcal{S}_{ij}^{[4q]}Tr\left[ A^{j}A^{i}\mathcal{P}^{2}%
\right]  \label{lttp2}
\end{eqnarray}%
where the dominant and the subdominant terms in the large-$N_{c}$ expansion
are considered and correspond to the decay diagrams expressed in Figs. 1.a
and 1.b , which are proportional to $c_{1}$ and $c_{2}$ respectively. In Eq.
(\ref{ltpp}) the interaction Lagrangian is expressed in terms of the diquark
matrices $D$ and $D^{\dag }$: in this way invariance under $SU_{V}(3),$ C
and P transformation is easily verified by using the transformation
properties listed in Table 1 and Table 2. In the form (\ref{lttp2}) the
tetraquark states are made explicit by using Eq. (\ref{s4q}): the decay
amplitudes for the tetraquark states into pseudoscalar mesons can be easily
evaluated from Eq. (\ref{lttp2}), see Ref. \cite{tq}.

Identifying the light scalar mesons as tetraquark states means the following
assignment \cite{tq}:%
\begin{equation}
\mathcal{S}^{[4q]}\equiv \left( 
\begin{array}{ccc}
\sqrt{\frac{1}{2}}(f_{B}-a_{0}^{0}(980)) & -a_{0}^{+}(980) & k^{+} \\ 
-a_{0}^{-}(980) & \sqrt{\frac{1}{2}}(f_{B}+a_{0}^{0}(980)) & -k^{0} \\ 
k^{-} & -\overline{k}^{0} & \sigma _{B}%
\end{array}%
\right)  \label{s4q2}
\end{equation}%
where $a_{0}[4q]$ and $k[4q]$ of Eqs. (\ref{tet1})-(\ref{tet2}) are
identified with the physical resonances $a_{0}(980)$ and $k(800).$ Then, a
mixing of the isoscalar tetraquark states $\sigma _{B}[4q]\equiv \sigma _{B}$
and $f_{B}[4q]\equiv f_{B}$, leading to the physical states $f_{0}(600)$ and 
$f_{0}(980)$, occurs \cite{tq}. The nonet $\mathcal{S}^{[4q]}$ transforms as
a usual scalar nonet under flavour, parity and charge transformations: $%
\mathcal{S}^{[4q]}\rightarrow U\mathcal{S}^{[4q]}U^{\dagger }$ ($U\subset
SU_{V}(3)$), $\mathcal{S}^{[4q]}$ and $\left( \mathcal{S}^{[4q]}\right) ^{t}$
respectively.

The assignment of Eq. (\ref{s4q2}), i.e. the interpretation of the light
scalar states as tetraquark resonances, has some characteristics able to
explain some enigmatic properties of the light scalar mesons: the almost
mass degeneracy of the state $a_{0}(980)$ and $f_{0}(980)$ and the strong
decay rates into $\overline{K}K$ are an immediate consequence of the quark
content of such states in this scenario. Then, in the analysis of \cite{tq}
the strengths of diagrams of Fig 1.a and Fig 1.b are analyzed
quantitatively: it has been found that the a sizable contribution of the
subdominant decay mechanism (Fig. 1.b), resulting in the ratio $%
c_{2}/c_{1}=0.62$, improves the theoretical prediction of the important
branching ratio $g_{f_{0}\rightarrow \overline{K}K}^{2}/g_{a_{0}\rightarrow 
\overline{K}K}^{2}$ \cite{bugg}. In fact, at the leading order
(OZI-superallowed, Fig. 1.a, $c_{2}=0$) one has $\left\vert
g_{f_{0}\rightarrow \overline{K}K}^{2}/g_{a_{0}\rightarrow \overline{K}%
K}^{2}\right\vert \leq 1,$ in clear contrast with the result $%
g_{f_{0}\rightarrow \overline{K}K}^{2}/g_{a_{0}\rightarrow \overline{K}%
K}^{2}=2.15\pm 0.40$ reported in the analysis of Refs. \cite{bugg,buggexp06}.

In the Lagrangian (\ref{ltpp}) only flavour symmetry, and not chiral
symmetry, is present. The basic question which we address in the present
work is what happens when extending the symmetry group. As we shall see, we
obtain mixing of the tetraquark and quarkonia scalar states. That is, the
strict equivalence of Eq. (\ref{s4q2}) is not anymore valid: the physical
scalar resonances below and above $1$ GeV will be an admixture of four-quark
and $\overline{q}q$ configurations. One crucial question is if the
tetraquark content for the light scalar states below 1 GeV (and
correspondingly quarkonia above 1 GeV) is the dominant one or not.

In order to see these phenomena at work we first consider the chiral partner
of the scalar diquark of Eq. (\ref{d}), a necessary step in order to write
down a chiral invariant interaction Lagrangian.

\subsection{ Pseudoscalar diquark}

The pseudoscalar diquark is the chiral partner of the scalar diquark and is
described by the diquark-matrix $\widetilde{D}$ and by the currents $%
\widetilde{\varphi }_{i}$:%
\begin{equation}
\widetilde{D}_{ij}\equiv \sqrt{\frac{1}{2}}\left(
q_{j}^{t}Cq_{i}-q_{i}^{t}Cq_{j}\right) =\sum_{i=1}^{3}\widetilde{\varphi }%
_{i}A^{i};\text{ }\widetilde{\varphi }_{i}=\sqrt{\frac{1}{2}}\varepsilon
_{ijk}q_{j}^{t}Cq_{k}\text{ .}  \label{dtilde}
\end{equation}%
The pseudoscalar diquark has the same flavor (and color) substructure ($%
\overline{3}_{F}$ ,$\overline{3}_{C}$) as the scalar diquark but negative
parity. It corresponds to:%
\begin{equation}
\left\vert qq\right\rangle _{L=S=1}=\left\vert \text{space: }%
L=1\right\rangle \left\vert \text{spin: }S=1\right\rangle \left\vert \text{%
color: }\overline{3}_{C}\right\rangle \left\vert \text{flavor: }\overline{3}%
_{F}\right\rangle ;\text{ }J^{P}=0^{-}\text{ .}  \label{diqmeno}
\end{equation}

The matrix $\widetilde{D}$ and the pseudoscalar diquarks $\widetilde{\varphi 
}_{i}$ transforms exactly as in Table 2 but with opposite parity.

In the chiral limit the scalar and the pseudoscalar diquarks have the same
mass. However, chiral symmetry is spontaneously broken by the QCD vacuum.
Calculations based on instantons show that a strong attraction is generated
in the scalar channel and a strong repulsion in the pseudoscalar one \cite%
{inst1,inst2}. Support for this picture is found in the recent Lattice
calculation of Ref. \cite{degrand}, in the chiral model for diquarks of Ref. 
\cite{diqeff}, in which the pseudoscalar diquark $\widetilde{D}$ is about
600 MeV heavier than the scalar partner, and in the framework of
Dyson-Schwinger equation \cite{maris}, where the mass difference is of the
same order of magnitude.

The common result of the above cited works is that the pseudoscalar diquark
is loosely bound and heavier when compared to the scalar partner. Indeed, it
is not clear if the pseudoscalar diquark can play the role of a constituent
for hadronic states. As emphasized in Ref. \cite{witten}, in the large-$%
N_{c} $ limit only quarkonia states survive in the mesonic sector, a fact
which also explain why non-quarkonia states are rare in the mesonic
spectrum. The scalar diquark, being the most compact diquark state, can
represent an exception and play a role in the physical world at $N_{c}=3.$
For all these reasons we will consider only the scalar diquark, and not the
pseudoscalar diquark, as a basic and compact constituent of low-energy
physical resonances. The inclusion of the pseudoscalar diquark is however a
necessary intermediate step in order to write down a chiral invariant
Lagrangian, see below.

\subsection{Chiral invariant interaction Lagrangian}

Out of the scalar and pseudoscalar matrices $D$ and $\widetilde{D}$ of Eqs. (%
\ref{d}) and (\ref{dtilde}) we define the matrices $D_{R}$ and $D_{L}$:%
\begin{equation}
D_{R}=\sqrt{\frac{1}{2}}(\widetilde{D}+D)=\sum_{i=1}^{3}\varphi
_{i}^{R}A^{i};\text{ }\varphi _{i}^{R}=\sqrt{\frac{1}{2}}\left( \widetilde{%
\varphi }_{i}+\varphi _{i}\right) ,
\end{equation}%
\begin{equation}
D_{L}=\sqrt{\frac{1}{2}}(\widetilde{D}-D)=\sum_{i=1}^{3}\varphi
_{i}^{L}A^{i};\text{ }\varphi _{i}^{R}=\sqrt{\frac{1}{2}}\left( \widetilde{%
\varphi }_{i}-\varphi _{i}\right) .
\end{equation}%
The transformation properties of the matrices $D_{R}$ and $D_{L}$ are
summarized in Table 3.

\begin{center}
\textbf{Table 3: }Properties of the diquark matrices $D_{R}$ and $D_{L}$.

$%
\begin{tabular}{|c|c|c|}
\hline
& $D_{R}=\sum_{i=1}^{3}\varphi _{i}^{R}A^{i}$ & $D_{L}=\sum_{i=1}^{3}\varphi
_{i}^{L}A^{i}$ \\ \hline
Currents & $\text{ }\varphi _{i}^{R}=\varepsilon _{ijk}q_{j}^{t}CP_{R}q$ & $%
\varphi _{i}^{L}=\varepsilon _{ijk}q_{j}^{t}CP_{L}q_{k}$ \\ \hline
P & $-D_{L}$ & $-D_{R}$ \\ \hline
C & $D_{R}^{\dagger }$ & $D_{L}^{\dagger }$ \\ \hline
$SU_{V}(3)$ & $UD_{R}U^{t}$ & $UD_{L}U^{t}$ \\ \hline
$SU_{R}(3)\times $ $SU_{L}(3)$ & $RD_{R}R^{t}$ & $LD_{L}L^{t}$ \\ \hline
$U_{A}(1)$ ($q_{i}\rightarrow e^{i\nu \gamma _{5}}q_{i}$) & $e^{2i\nu }D_{R}$
& $e^{-2i\nu }D_{L}$ \\ \hline
\end{tabular}%
$
\end{center}

\bigskip

Under chiral transformations the diquark components $\varphi _{i}^{R}$
transform as a right-handed antiquark, while the components $\varphi
_{i}^{L} $ as a left-handed antiquark:%
\begin{equation}
\varphi _{i}^{R}\rightarrow \varphi _{k}^{R}R_{ki}^{\dagger },\text{ }%
\varphi _{i}^{L}\rightarrow \varphi _{k}^{L}L_{ki}^{\dagger }\text{ under }%
SU_{R}(3)\times SU_{L}(3).\text{ }
\end{equation}

We are now in the position to write a chiral invariant interaction
Lagrangian in terms of the diquark matrices $D_{R}$ and $D_{L}$ and the
quarkonia nonet matrix $\Sigma .$ By taking into account the transformation
properties in Table 1 and Table 3 the chiral invariant interaction
Lagrangian at leading and subleading order in the large-$N_{c}$ expansion
reads:%
\begin{equation}
\mathcal{L}_{c.i.}=-c_{1}Tr\left( D_{R}\Sigma ^{t}D_{L}^{\dagger }\Sigma
+D_{L}\Sigma ^{\ast }D_{R}^{\dagger }\Sigma ^{\dagger }\right)
+c_{2}Tr\left( D_{R}D_{R}^{\dagger }\Sigma ^{\dagger }\Sigma
+D_{L}D_{L}^{\dagger }\Sigma \Sigma ^{\dagger }\right) .  \label{lchirinv}
\end{equation}%
A diquark and an antidiquark matrices are coupled to two $\Sigma $'s: in
both cases two quarks and two antiquarks are present. The Lagrangian (\ref%
{lchirinv}) is also invariant under parity, charge conjugation and $U_{A}(1)$
axial transformations.

The constants $c_{1}$ and $c_{2}$ are exactly those of Eq. (\ref{ltpp}). In
fact, the flavor invariant Lagrangian (\ref{ltpp}) has to emerge out of the
chiral invariant Lagrangian. We discuss the precise relation between Eq. (%
\ref{ltpp}) and Eq. (\ref{lchirinv}) in the next section.

The presence of two different diquark types leads to 4 tetraquark nonets:
two scalars given by $\varphi _{i}^{\dagger }\varphi _{j}$ ($=\mathcal{S}%
^{[4q]}$) and $\widetilde{\varphi }_{i}^{\dagger }\widetilde{\varphi }_{j}$
and two pseudoscalars given by $\varphi _{i}^{\dagger }\widetilde{\varphi }%
_{j}$ and $\widetilde{\varphi }_{i}^{\dagger }\varphi _{j}$ (admixtures of
these nonets with definite properties under chiral transformations are
found, see Appendix A).

As discussed in Section 2.3 we do not consider the pseudoscalar diquark of
Eq. (\ref{diqmeno}) as a suitable constituent for mesonic states. For this
reason we consider only the scalar diquark as relevant constituent for
low-energy spectroscopy, thus only the tetraquark nonet $\mathcal{S}%
^{[4q]}=\varphi _{i}^{\dagger }\varphi _{j}$ is taken into account. The
other three nonets may eventually exist, but be heavier, and/or too broad to
be measured. In the QCD spectrum below 2 GeV one notices the presence of
supernumerary scalar states, which can accommodate a non-quarkonia nonet
like $\mathcal{S}^{[4q]}$ (and probably a scalar glueball), but the presence
of a second non-quarkonia scalar nonet, such as the composition of two
pseudoscalar diquarks $\widetilde{\varphi }_{i}^{\dagger }\widetilde{\varphi 
}_{j}$, seems to be excluded by present data \cite{pdg}.

The pseudoscalar sector is less clear: beyond the well established
low-energy pseudoscalar nonet $\{\pi ,K,\eta ,\eta ^{\prime }\}$, a second
nonet shows up at around $1.3$ GeV: the state $\pi (1300)$ is usually
interpreted as the radial excitation of the pion \cite{amslerrev}. A kaonic
state $K(1460)$ is also reported in \cite{pdg}. The two isoscalar states $%
\eta (1295)$ and $\eta (1475)$ are usually interpreted as the excited $\eta $
and $\eta ^{\prime }$ mesons. The resonance $\eta (1405)$ is ambiguous, and
various interpretations have been proposed, such as a pseudoscalar glueball,
but some authors do not accept its existence \cite{klempt}. Other massive
pseudoscalar states such as $\pi (1800),$ $K(1830),$ $\eta (1760)$ are
identified and interpreted as the second radial excitation \cite{amslerrev}
(but this assignment is not yet conclusive).

The fact that we take into account only scalar diquarks and the
corresponding scalar nonet $\mathcal{S}^{[4q]}=\varphi _{i}^{\dagger
}\varphi $ is the basic difference with Refs. \cite%
{black,fariborz,napsuciale,fariborzlast}, where a scalar and a pseudoscalar
nonets are considered (see also Appendix A). For instance, the resonance $%
\pi (1300)$ is mainly a four-quark states in Ref. \cite{fariborz}.
Furthermore, the Lagrangian interaction of Refs. \cite{black,napsuciale}
breaks $U_{A}(1)$ invariance, while Eq. (\ref{lchirinv}) does not. Here we
do not evaluate the masses of quarkonia (we did not specify the potential $%
V_{0}+V_{SB}$ in section 2.1) and of tetraquark states, but we concentrate
on their interaction. Theoretical evaluation of masses of bare states is, on
the contrary, an important part of Refs. \cite%
{black,fariborz,napsuciale,fariborzlast}.

\section{Light tetraquark states: mixing with scalar quarkonia and
condensates}

\subsection{The `remnant' interaction Lagrangian}

By isolating in the interaction Lagrangian (\ref{lchirinv}) only those terms
involving the scalar diquark matrix $D$ (and not the pseudoscalar matrix $%
\widetilde{D}$) we obtain:%
\begin{equation}
\mathcal{L}_{\mathcal{S}^{[4q]}\Sigma \Sigma }=\frac{c_{1}}{2}Tr\left[
D\Sigma ^{t}D^{\dagger }\Sigma +D\Sigma ^{\ast }D^{\dagger }\Sigma ^{\dagger
}\right] +\frac{c_{2}}{2}Tr\left[ DD^{\dagger }\Sigma ^{\dagger }\Sigma
+DD^{\dagger }\Sigma \Sigma ^{\dagger }\right]  \label{lagd}
\end{equation}%
\begin{equation}
=-\frac{c_{1}}{2}\mathcal{S}_{ij}^{[4q]}Tr\left[ A^{j}\Sigma ^{t}A^{i}\Sigma
+D\Sigma ^{\ast }D^{\dagger }\Sigma ^{\dagger }\right] -\frac{c_{2}}{2}%
\mathcal{S}_{ij}^{[4q]}Tr\left[ A^{j}A^{i}\Sigma ^{\dagger }\Sigma
+A^{j}A^{i}\Sigma \Sigma ^{\dagger }\right] \text{ ,}
\end{equation}%
where in the last line the expression is explicitly presented in terms of
the tetraquark scalar nonet $\mathcal{S}^{[4q]}$ defined in Eq. (\ref{s4q}).

For completeness we report the total Lagrangian under consideration. It is
the sum of the Lagrangian $\mathcal{L}_{\Sigma }$ in Eq. (\ref{lsigma}),
which involves the pseudoscalar and the scalar quarkonia nonets, of a
quadratic Lagrangian involving the kinematic and the mass terms of the
scalar tetraquark nonet $\mathcal{S}^{[4q]}$ and of the quarkonia-tetraquark
interaction $\mathcal{L}_{\mathcal{S}^{[4q]}\Sigma \Sigma }$ of Eq. (\ref%
{lagd}): 
\begin{equation}
\mathcal{L}_{\text{tot}}=\mathcal{L}_{\Sigma }+\mathcal{L}_{\mathcal{S}%
^{[4q]}\text{-quadratic}}+\mathcal{L}_{\mathcal{S}^{[4q]}\Sigma \Sigma }%
\text{ .}  \label{ltot}
\end{equation}%
The term $\mathcal{L}_{\mathcal{S}^{[4q]}\text{-quadratic}}$ is described in
Ref. \cite{tq}, where the nonet mass splitting and the isoscalar-mixing are
taken into account. In the present work we do not need to specify it. Our
attention is focused on the quarkonia-tetraquark interaction term $\mathcal{L%
}_{\mathcal{S}^{[4q]}\Sigma \Sigma }$.

The phenomenon of chiral symmetry breaking, encoded in the non-vanishing $%
vev $ for the field $\Sigma $ in Eq. (\ref{csb1}), introduces further terms
beyond the tetraquark-quarkonia decay diagrams of Fig.1: a mixing term among
the two scalar nonets $\mathcal{S}^{[4q]}$ and $\mathcal{S}$ and a linear
term in $\mathcal{S}^{[4q]},$ corresponding to non-vanishing tetraquark
condensates, are generated. In fact, when substituting $\Sigma =\Sigma _{0}+%
\mathcal{S}+i\mathcal{P}$ into (\ref{lagd}), we can decompose it into four
different terms:%
\begin{equation}
\mathcal{L}_{\mathcal{S}^{[4q]}\Sigma \Sigma }=\mathcal{L}_{\mathcal{S}%
^{[4q]}PP}+\mathcal{L}_{\mathcal{S}^{[4q]}SS}+\mathcal{L}_{\text{mix}}+%
\mathcal{L}_{4q\text{-cond}}  \label{decom}
\end{equation}

The Lagrangian $\mathcal{L}_{\mathcal{S}^{[4q]}PP}$ of Eq. (\ref{ltpp}) is
reobtained (with the same coupling strengths $c_{1}$ and $c_{2}$). The
Lagrangian $\mathcal{L}_{\mathcal{S}^{[4q]}SS}$ is analogous to $\mathcal{L}%
_{\mathcal{S}^{[4q]}PP}$, where one has two scalar quarkonia mesons instead
of two pseudoscalar ones. We will not study the phenomenological
implications of this term because in the present work the bare tetraquark
states are lighter than the quarkonia states, thus such a decay is not
kinematically allowed.

The term $\mathcal{L}_{\text{mix}}$ is linear in $\Sigma _{0}$ and describes
the mixing of $\mathcal{S}^{[4q]}$ and $\mathcal{S}$, see next subsection.
The term $\mathcal{L}_{4q\text{-cond}}$ is quadratic in $\Sigma _{0}$ and
linear in $\mathcal{S}^{[4q]}$ and is responsible for non-zero vacuum
expectation value of the the isoscalar tetraquark fields (see Section 3.3).

In Eq. (\ref{decom}) we obtained a decomposition of the tetraquark
interaction terms by expanding $\Sigma $ around $\Sigma _{0}$, which is a
minimum for the potential $V_{0}+V_{SB}$ as discussed in Section 2.1. Care
is however needed: when including the interaction term $\mathcal{L}_{%
\mathcal{S}^{[4q]}\Sigma \Sigma }$ of Eq. (\ref{lagd}) in the total
Lagrangian $\mathcal{L}_{\text{tot}}$ of Eq. (\ref{ltot}) the potential
involving $\Sigma $-matrix has been extended to $V_{0}+V_{SB}-\mathcal{L}_{%
\mathcal{S}^{[4q]}\Sigma \Sigma }$: the matrix $\Sigma _{0}$ is not anymore
the minimum. Strictly speaking one should not expand around $\Sigma _{0}$ as
in Eq. (\ref{decom}) but around the new minimum of $V_{0}+V_{SB}-\mathcal{L}%
_{\mathcal{S}^{[4q]}\Sigma \Sigma }$. We will discuss the issue in detail in
Section 3.3, where we show that in our case $\Sigma _{0}$ still represents a
good approximation for the minimum and that the expansion of Eq. (\ref{decom}%
) is justified. We also illustrate the point by means of a simple toy-model.

\subsection{Scalar tetraquark-quarkonia mixing in the isovector sector}

\subsubsection{The mixing Lagrangian}

The tetraquark-quarkonia mixing Lagrangian $\mathcal{L}_{mix}$ is derived
from Eq. (\ref{lagd}) by using $\Sigma =\Sigma _{0}+\mathcal{S}+i\mathcal{P}$
with $\Sigma _{0}=diag\{\alpha _{u},\alpha _{u},\alpha _{s}\}$ and keeping
terms linear in $\Sigma _{0}$: 
\begin{equation*}
\mathcal{L}_{\text{mix}}=c_{1}Tr\left[ D\Sigma _{0}D^{\dagger }\mathcal{S}+D%
\mathcal{S}^{t}D^{\dagger }\Sigma _{0}\right] +c_{2}Tr\left[ DD^{\dagger
}\Sigma _{0}\mathcal{S}+DD^{\dagger }\mathcal{S}\Sigma _{0}\right]
\end{equation*}%
\begin{equation}
=-c_{1}\mathcal{S}_{ij}^{[4q]}Tr\left[ A^{j}\Sigma _{0}A^{i}\mathcal{S}+A^{j}%
\mathcal{S}^{t}A^{i}\Sigma _{0}\right] -c_{2}\mathcal{S}_{ij}^{[4q]}Tr\left[
A^{j}A^{i}(\Sigma _{0}\mathcal{S}+\mathcal{S}\Sigma _{0})\right]
\label{lmix}
\end{equation}%
We depict the process corresponding to $\mathcal{L}_{mix}$ in Fig. 2, where
the two diagrams resemble Figs 1.a and 1.b, but at one vertex the vacuum
expectation matrix $\Sigma _{0}$ enters in the game. If $\Sigma _{0}$
vanishes, such terms vanish as well. It is noticeable that the
decay-strengths parameters $c_{1}$ and $c_{2}$ also regulate the intensity
of the mixing. In Eq. (\ref{s4q2}) and in Refs. \cite{jaffe,maiani,tq} the
scalar states below 1 GeV are interpreted as pure tetraquark states. The
present analysis shows that such an assignment cannot be strictly valid
because mixing occurs. We aim now to evaluate the intensity of this mixing
in the isovector channel.

An important point is the following: in section 2.1 we discussed various
arguments in favour of bare quarkonia masses well above $1$ GeV. At the same
time in the Introduction and in Section 2.4 we recalled that the scalar
diquark emerges a compact light object within different approaches
(one-gluon exchange \cite{amslerrev,onegluon}, instantons \cite{inst1,inst2}%
, NJL model and DSE \cite{weise,maris}). The s-wave tetraquark states
arising by composition of a diquark and antidiquark (as expressed in Eq. (%
\ref{s4q}) and (\ref{tet1})-(\ref{tet2})) is expected to have a mass below
(or about) 1 GeV, as discussed in Sections 2.1 and 2.2 by means of
phenomenological arguments and as suggested by the Lattice works of Refs. 
\cite{jaffelatt,okiharu,mathur}. These facts lead us to consider the bare
level ordering $M_{4q}<M_{\overline{q}q}.$

\subsubsection{Mixing in the isovector channel}

We analyze the mixing of the two neutral $a_{0}^{0}$ states, denoted as $%
a_{0}^{0}[4q]$ (from the tetraquark nonet $\mathcal{S}^{[4q]}$ of Eq. (\ref%
{tet1})-(\ref{tet2})) and as $a_{0}^{0}[\overline{q}q]$ (from the quarkonia
nonet $\mathcal{S}$ of Table 1). The isovector channel is free from
isoscalar-mixing (and glueball) complications, and is experimentally better
known than the kaonic sector.

We isolate in $\mathcal{L}_{mix}$ of Eq. (\ref{lmix}) the part concerning
the neutral $a_{0}$ states:%
\begin{equation}
\mathcal{L}_{\text{mix-}a_{0}^{0}}=2(c_{1}\alpha _{s}+c_{2}\alpha
_{u})\left( a_{0}^{0}[4q]\cdot a_{0}^{0}[\overline{q}q]\right) .
\label{lmixa0}
\end{equation}%
When including the kinematic and (bare) mass contributions one has to
diagonalize the following Lagrangian:%
\begin{equation}
\mathcal{L=}\frac{1}{2}\left( \partial _{\mu }a_{0}^{0}[4q]\right) ^{2}+%
\frac{1}{2}\left( \partial _{\mu }a_{0}^{0}[\overline{q}q]\right) ^{2}+\frac{%
1}{2}v^{t}\Omega v
\end{equation}%
where%
\begin{equation}
v=\left( 
\begin{array}{c}
a_{0}^{0}[4q] \\ 
a_{0}^{0}[\overline{q}q]%
\end{array}%
\right) ,\text{ }\Omega =\left( 
\begin{array}{cc}
-M_{a_{0}^{0}[4q]}^{2} & 2(c_{1}\alpha _{s}+c_{2}\alpha _{u}) \\ 
2(c_{1}\alpha _{s}+c_{2}\alpha _{u}) & -M_{a_{0}^{0}[\overline{q}q]}^{2}%
\end{array}%
\right) .
\end{equation}%
The orthogonal transformation matrix $B$, given by 
\begin{equation}
B\Omega B^{t}=-diag\{M_{a_{0}(980)}^{2},M_{a_{0}(1470)}^{2}\}\text{ ,}
\label{diagmass}
\end{equation}%
connects the bare tetraquark and quarkonia states to the physical ones:%
\begin{equation}
\left( 
\begin{array}{c}
a_{0}^{0}(980) \\ 
a_{0}^{0}(1470)%
\end{array}%
\right) =B\cdot \left( 
\begin{array}{c}
a_{0}^{0}[4q] \\ 
a_{0}^{0}[\overline{q}q]%
\end{array}%
\right) ,\text{ }B=\left( 
\begin{array}{cc}
\cos (\theta ) & \sin (\theta ) \\ 
-\sin (\theta ) & \cos (\theta )%
\end{array}%
\right) .  \label{mixing}
\end{equation}

The physical masses read \cite{pdg}:

\begin{equation}
M_{a_{0}(980)}=984.7\pm 1.2\text{ MeV and }M_{a_{0}(1450)}=1474\pm 19\text{
MeV.}  \label{physmass}
\end{equation}

The decay rates for the decay channel $a_{0}(980)\rightarrow \eta \pi $ and $%
a_{0}(1450)\rightarrow \eta \pi $ are given by:%
\begin{equation}
\Gamma _{a_{0}(980)\rightarrow \eta \pi }=\frac{p_{a_{0}(980)\rightarrow
\eta \pi }}{8\pi M_{a_{0}(980)}^{2}}g_{_{a_{0}(980)\rightarrow \eta \pi
}}^{2},\text{ }\Gamma _{a_{0}(1450)\rightarrow \eta \pi }=\frac{%
p_{a_{0}(1450)\rightarrow \eta \pi }}{8\pi M_{a_{0}(1450)}^{2}}%
g_{_{a_{0}(1450)\rightarrow \eta \pi }}^{2},
\end{equation}%
where $p_{a_{0}(980)\rightarrow \eta \pi }$ and $p_{a_{0}(1450)\rightarrow
\eta \pi }$ represent the phase-space factors and the decay amplitudes $%
g_{_{a_{0}(980)\rightarrow \eta \pi }}$ and $g_{_{a_{0}(1450)\rightarrow
\eta \pi }}$ are a superposition of the tetraquark and quarkonia
contributions:%
\begin{eqnarray}
g_{_{a_{0}(980)\rightarrow \eta \pi }}^{2} &=&\left[ g_{_{a_{0}[4q]%
\rightarrow \eta \pi }}\cos (\theta )+g_{_{a_{0}[\overline{q}q]\rightarrow
\eta \pi }}\sin (\theta )\right] ^{2},  \notag \\
g_{_{a_{0}(1450)\rightarrow \eta \pi }}^{2} &=&\left[ -g_{_{a_{0}[4q]%
\rightarrow \eta \pi }}\sin (\theta )+g_{_{a_{0}[\overline{q}q]\rightarrow
\eta \pi }}\cos (\theta )\right] ^{2}.  \label{g}
\end{eqnarray}%
The amplitude $g_{_{a_{0}[4q]\rightarrow \eta \pi }}$ is calculated from $%
\mathcal{L}_{\mathcal{S}^{[4q]}PP}$ of Eq. (\ref{ltpp}) and reads \cite{tq}:%
\begin{equation}
g_{_{a_{0}[4q]\rightarrow \eta \pi }}=\frac{2c_{1}}{\sqrt{3}}\left[ \sqrt{2}%
\cos (\theta _{P})+\sin (\theta _{P})\right] +\sqrt{\frac{2}{3}}c_{2}\left[
\cos (\theta _{P})-\sqrt{2}\sin (\theta _{P})\right]  \label{ga04qtheor}
\end{equation}%
where $\theta _{P}=-9.95^{\circ }$ at tree-level \cite%
{giacosachiral,venugopal}.

The quantity $g_{_{a_{0}[\overline{q}q]\rightarrow \eta \pi }}$ depends on
the Lagrangian describing the decay of scalar quarkonia into pseudoscalar
mesons, which we did not specify in this work. In the following we will
treat it as a free parameter, see below.

We now turn the attention to the experimental informations about the
coupling constants in Eq. (\ref{ga04qtheor}). The coupling constant $%
g_{_{a_{0}(980)\rightarrow \eta \pi }}^{2}$ as extracted from experimental
analyses varies between $5$ and $10$ GeV$^{2}$ \cite{flatte}. We then
consider the following three values in the above range\footnote{%
Because of the large uncertanties in the experimental analyses we do not
report in Eq. (\ref{gqa0980}) a single value with corresponding errors, but
three possible values in agreement with present experimental infromations.}:%
\begin{equation}
g_{_{a_{0}(980)\rightarrow \eta \pi }}^{2}=5,\text{ }7.5,\text{ }10\text{ GeV%
}^{2}  \label{gqa0980}
\end{equation}%
which corresponds to $\Gamma _{a_{0}(980)\rightarrow \eta \pi }=65.8,$ $98.7$%
, $131.6$ MeV respectively. These values are compatible with the data
reported by PDG \cite{pdg}, which are however not yet precise. By using%
\begin{equation*}
\Gamma _{a_{0}(980)}=\Gamma _{a_{0}(980)\rightarrow \overline{K}K}+\Gamma
_{a_{0}(980)\rightarrow \eta \pi }\text{ and }\frac{\Gamma
_{a_{0}(980)\rightarrow \overline{K}K}}{\Gamma _{a_{0}(980)\rightarrow \eta
\pi }}=0.183\pm 0.024\text{ \cite{pdg},}
\end{equation*}%
we obtain (ignoring the error on the last ratio): $\Gamma
_{a_{0}(980)\rightarrow \eta \pi }=\Gamma _{a_{0}}/1.183.$ In PDG \cite{pdg}
the value $\Gamma _{a_{0}(980)}=50$-$100$ MeV is reported, thus implying $%
\Gamma _{a_{0}(980)\rightarrow \eta \pi }$ equal to $42.3$ and $84.6$ MeV
respectively. The largest value $g_{_{a_{0}(980)\rightarrow \eta \pi
}}^{2}=10$ GeV$^{2}$ seems disfavored and we regard it as an upper limit.

We now turn the attention to $g_{_{a_{0}(1450)\rightarrow \eta \pi }}^{2}.$
In Ref. \cite{pdg} the averages for the following branching ratios are
reported: 
\begin{equation*}
\frac{\Gamma _{a_{0}(1450)\rightarrow \eta ^{\prime }\pi }}{\Gamma
_{a_{0}(1450)\rightarrow \eta \pi }}=0.35\pm 0.16,\text{ }\frac{\Gamma
_{a_{0}(1450)\rightarrow \overline{K}K}}{\Gamma _{a_{0}(1450)\rightarrow
\eta \pi }}=0.88\pm 0.23.
\end{equation*}%
The full width amounts to $\Gamma _{a_{0}(1450)}=265\pm 13$ MeV. The
contribution of the two-pseudoscalar decays to the full width is unknown. By
assuming it to be dominant, and thus that the $\omega \rho $ mode is
suppressed, we obtain $\Gamma _{a_{0}(1450)\rightarrow \eta \pi }\simeq 119$
MeV, corresponding to $g_{_{a_{0}(1450)\rightarrow \eta \pi }}^{2}{}\simeq
10.4$ GeV$^{2}.$ We do not include errors because we ignore the contribution
of the $\omega \rho $ decay to the full width. Furthermore, the experimental
result $\Gamma _{a_{0}(1450)\rightarrow \omega \rho }/\Gamma
_{a_{0}(1450)\rightarrow \eta \pi }=10.7\pm 2.3$ reported in Ref. \cite%
{baker} would indicate a dominant $\omega \rho $ mode. This value is however
not listed as an average or fit in \cite{pdg}. We will consider the value 
\begin{equation}
g_{_{a_{0}(1450)\rightarrow \eta \pi }}^{2}=10.4\text{ GeV}^{2}\text{,}
\label{gqa01450}
\end{equation}%
being aware that it could be smaller.

We now turn to the evaluation of the mixing angle. We consider the
theoretical coupling $g_{_{a_{0}[\overline{q}q]\rightarrow \eta \pi }}$ of
Eq. (\ref{g}) as a free parameter, thus we are left with five parameters: $%
\{c_{1},c_{2},M_{a_{0}^{0}[4q]},M_{a_{0}^{0}[\overline{q}q]},g_{_{a_{0}[%
\overline{q}q]\rightarrow \eta \pi }}\}.$ We fix the ratio $c_{2}/c_{1}=0.62$
as obtained in \cite{tq}, where the light scalar are interpreted as
tetraquark states.\ Although this choice cannot be a priori justified, we
will then vary the ratio $c_{2}/c_{1}$ checking the dependence of the
results on it.

We fix the remaining four parameters to the physical masses of Eq. (\ref%
{physmass}), to the intermediate value $g_{_{a_{0}(980)\rightarrow \eta \pi
}}^{2}=7.5$ GeV$^{2}$ of Eq. (\ref{gqa0980}) and to Eq. (\ref{gqa01450}).

By using the bare level ordering $M_{a_{0}^{0}[4q]}<M_{a_{0}^{0}[\overline{q}%
q]}$ the parameters are determined as (values in GeV, ratio $%
c_{2}/c_{1}=0.62 $ fixed):

\begin{equation}
\left\vert c_{1}\right\vert =0.96\text{, }M_{a_{0}^{0}[4q]}=1.01,\text{ }%
M_{a_{0}^{0}[\overline{q}q]}=1.45,\text{ }g_{_{a_{0}[\overline{q}%
q]\rightarrow \eta \pi }}=3.75\text{ .}  \label{solfulls}
\end{equation}%
corresponding to a quarkonium amount in the resonance $a_{0}(980)$: 
\begin{equation}
(\sin \theta )^{2}=4.93\%\text{ }.
\end{equation}

According to our result the resonance $a_{0}(980)$ has a by far dominant
tetraquark substructure and only a small quarkonium amount. Similarly, the
resonance $a_{0}(1450)$ has a dominant quarkonium substructure with a small $%
4.93\%$ tetraquark content. The mixing between the tetraquark and quarkonia
states turns out to be small.

To have used $c_{2}/c_{1}=0.62$ from \cite{tq} is then justified a
posteriori. Anyway, when varying the ratio $c_{2}/c_{1}$ and the couplings
of Eqs. (\ref{gqa0980})-(\ref{gqa01450}) the results show a stable behavior:
the mixing turns out to be small for all reasonable parameter choices, see
Appendix B.

Notice that we cannot determine the sign of the mixing angle $\theta $ and $%
c_{1}.$ In fact, we have no information about the sign of $%
g_{_{a_{0}(1450)\rightarrow \eta \pi }}$ and $g_{_{a_{0}(980)\rightarrow
\eta \pi }}$ from experiment. For this reason the modulus $\left\vert
c_{1}\right\vert $ is reported in Eq. (\ref{solfulls}). If $c_{1}>0,$ then $%
\theta >0$ and vice-versa. The two possibilities are however
indistinguishable here.

\subsubsection{ Further discussion}

Some comments are in other:

a) In the kaonic sector the situation is similar. For instance, the part of
the Lagrangian $\mathcal{L}_{mix}$ (\ref{lmix}) describing the $k^{-}[4q]$-$%
K_{0}^{-}[\overline{q}q]$ mixing reads:

\begin{equation}
\mathcal{L}_{\text{mix-}k}=-(2c_{1}\alpha _{u}+c_{2}(\alpha _{u}+\alpha
_{s}))\left( k^{-}[4q]\cdot K_{0}^{+}[\overline{q}q]\right) .
\end{equation}

By using the solution reported in Eq. (\ref{solfulls}) and the masses $%
M_{k(800)}\simeq 800$ MeV and $M_{K_{0}(1430)}=1414\pm 6$ MeV we obtain a
quarkonium amount in $k(800)$ of the order of $3\%,$ i.e. very small. As a
consequence, the state $K_{0}(1430)$ has a dominant quarkonium content. The
corresponding bare masses of the tetraquark and quarkonia states are $%
M_{k[4q]}=823$ MeV and $M_{K[\overline{q}q]}=1400$ MeV, thus only slightly
shifted from the physical masses.

Let us turn to the problematic $\pi K$-decay of the two resonances: the
smallness of the mixing allows us to consider the approximate relations $%
g_{k(800)\rightarrow \pi K}\simeq g_{k[4q]\rightarrow \pi K}$ ($=\sqrt{3}(%
\sqrt{2}c_{1}+c_{2}/\sqrt{2})$ as evaluated in \cite{tq}) and $%
g_{K(1430)\rightarrow \pi K}\simeq g_{K[\overline{q}q]\rightarrow \pi K}.$
Using the parameters of (\ref{solfulls}) one finds $\Gamma
_{k(800)\rightarrow \pi K}\simeq 130$ MeV. The uncertainty on the decay
strengths $c_{1}$ and $c_{2}$ allows for a width between $100$ and $220$
MeV. The values in this range are smaller than the present (not yet
conclusive) experimental value of about $600$ MeV \cite{pdg}. We refer to
the discussions about the experimental caveats in Ref. \cite{bugg}, where it
is also pointed out that quarkonium, molecular or tetraquark interpretations
all fail in reproducing the large width of $k(800)$. Meson-meson interaction
governed by chiral symmetry as presented in Ref. \cite{oller} can play an
important role to explain the large width of $k(800)$. It is important to
stress that also the decay $K(1430)\rightarrow \pi K$, evaluated in Ref. 
\cite{giacosachiral}, turns out to be smaller than the present data of a
factor 4. This problem has been analyzed in detail in Ref. \cite%
{giacosachiral} and is rather model-independent. Further work both on
theoretical and experimental sides is clearly needed to fit all the
properties of the kaonic states $k(800)$ and $K_{0}(1430)$ in a unified
picture.

b) We report the mixing Lagrangian in the isoscalar sector in terms of the
bare states $\sigma _{B}[4q]=\frac{1}{2}[u,d][\overline{u},\overline{d}]$, $%
f_{B}[4q]=\frac{1}{2\sqrt{2}}([u,s][\overline{u},\overline{s}]+[d,s][%
\overline{d},\overline{s}])$ and $N=\sqrt{\frac{1}{2}}(\overline{u}u+%
\overline{d}d),$ $S=\overline{s}s$ : 
\begin{eqnarray}
\mathcal{L}_{\text{mix-iso}} &=&2(c_{1}\alpha _{s}+c_{2}\alpha
_{u})Nf_{B}[4q]+2\sqrt{2}(c_{1}\alpha _{u}+c_{2}\alpha _{s})Sf_{B}[4q] 
\notag \\
&&+2\sqrt{2}(c_{1}+c_{2})\alpha _{u}N\sigma _{B}[4q].  \label{lmixiso}
\end{eqnarray}

The intensity of the mixing is of the same order of magnitude of the
isovector and isodoublet channel, that is small. The system is then
complicated by internal mixing terms like $\sigma _{B}f_{B}$ and $NS$ and
glueball mixing, which lead to the resonances $f_{0}(600)$ and $f_{0}(980)$
below 1 GeV, and to $f_{0}(1370),$ $f_{0}(1500)$ and $f_{0}(1710)$ above 1
GeV. Here we do not analyze this system quantitatively (see Ref. \cite%
{fariborzlast} for such a study). However, the tetraquark-quarkonia mixing
is small as verified in the isovector and isodoublet channels, thus it is
still valid to deal with two separated tetraquark and quarkonia nonets with
the scalar glueball intruding in the scalar-isoscalar quarkonia sector
between 1 and 2 GeV. We therefore expect smaller mixing than in Ref. \cite%
{fariborzlast}.

c) We did not take into account the momentum-dependence of the theoretical
amplitudes $g_{_{a_{0}[\overline{q}q]\rightarrow \eta \pi }}$ and $%
g_{_{a_{0}[4q]\rightarrow \eta \pi }}.$ For instance, within a chiral
perturbation theory framework the quantity $g_{_{a_{0}[\overline{q}%
q]\rightarrow \eta \pi }}$ has a (dominant) momentum-squared $p^{2}$
dependence of the form $g_{_{a_{0}[\overline{q}q]\rightarrow \eta \pi
}}\propto (p^{2}-M_{\pi }^{2}-M_{\eta }^{2})$ \cite{giacosachiral}. When
including such form in the calculation the results are similar. In general
(reasonable) momentum dependence does not change the picture, see Appendix B.

d) One of the basic starting points of the evaluation of the mixing has been
the bare level ordering $M_{a_{0}^{0}[4q]}<M_{a_{0}^{0}[\overline{q}q]}.$
The reasons for this choice have been listed in Sections 2.1 and 3.2.1. Here
we notice that solutions are possible also for the reversed bare level
ordering $M_{a_{0}^{0}[4q]}>M_{a_{0}^{0}[\overline{q}q]},$ for which the
quarkonium content is dominant in $a_{0}(980).$ Although the case $%
M_{a_{0}^{0}[4q]}>M_{a_{0}^{0}[\overline{q}q]}$ seems unlikely for the above
mentioned discussions, it cannot be ruled out.

e) The interaction Lagrangian $\mathcal{L}_{c.i.}$ of Eq. (\ref{lchirinv})
contains the dominant and subdominant terms in large-$N_{c}$ expansion.\
Further large-$N_{c}$ suppressed terms and flavor-symmetry breaking terms
were not included in the present analysis. Although they can quantitatively
influence the results, they are not believed to change the qualitative
picture emerging from this work.

\subsection{Tetraquark condensates}

\subsubsection{Vacuum expectation values and estimation of tetraquark
condensates}

The term $\mathcal{L}_{4q\text{-}cond}$ of Eq. (\ref{decom}) is linear in $%
\mathcal{S}^{[4q]}$ and quadratic in $\Sigma _{0}$. It explicitly reads:%
\begin{eqnarray}
\mathcal{L}_{4q\text{-cond}} &=&c_{1}Tr\left[ D\Sigma _{0}D^{\dagger }\Sigma
_{0}\right] +c_{2}Tr\left[ DD^{\dagger }\Sigma _{0}^{2}\right]  \notag \\
&=&2(c_{1}+c_{2})\alpha _{u}^{2}\sigma _{B}[4q]+\sqrt{2}\left( 2c_{1}\alpha
_{u}\alpha _{s}+c_{2}(\alpha _{u}^{2}+\alpha _{s}^{2})\right) f_{B}[4q]
\label{lconds}
\end{eqnarray}%
where in the second line the flavor-trace has been performed and a linear
dependence in the isoscalar fields $\sigma _{B}[4q]$ and $f_{B}[4q]$ is
found. This implies non-zero vacuum expectation values for these two (bare)
fields: 
\begin{equation}
\left\langle \sigma _{B}[4q]\right\rangle =\frac{2(c_{1}+c_{2})\alpha
_{u}^{2}}{M_{\sigma _{B[4q]}}^{2}},\text{ }\left\langle
f_{B}[4q]\right\rangle =\sqrt{2}\frac{2c_{1}\alpha _{u}\alpha
_{s}+c_{2}(\alpha _{u}^{2}+\alpha _{s}^{2})}{M_{f_{B}[4q]}^{2}},
\label{zeroval}
\end{equation}%
where $M_{\sigma _{B}[4q]}$ and $M_{f_{B}[4q]}^{2}$ refer to the (bare)
tetraquark masses of the states $\sigma _{B}[4q]$ and $f_{B}[4q].$

The non-zero vacuum expectation value ($vev$) of the nonet $\mathcal{S}%
^{[4q]}$ (\ref{tet1})-(\ref{tet2}) reads:%
\begin{equation}
\left\langle \mathcal{S}_{ij}^{[4q]}\right\rangle =\beta _{i}\delta _{ij},%
\text{ }\beta _{1}=\beta _{2}=\frac{\left\langle f_{B}[4q]\right\rangle }{%
\sqrt{2}},\text{ }\beta _{3}=\left\langle \sigma _{B}[4q]\right\rangle .
\label{tetvac}
\end{equation}%
We can estimate the corresponding tetraquark condensate following the
discussion of \cite{fariborz}:%
\begin{equation}
\left\vert \left\langle \frac{1}{2}\left( \varepsilon _{ikl}\overline{q}%
_{k}^{t}C\gamma ^{5}\overline{q}_{l}\right) \left( \varepsilon
_{jrm}q_{r}^{t}C\gamma ^{5}q_{m}\right) \right\rangle \right\vert \sim
\Lambda _{QCD}^{5}\beta _{i}\delta _{ij}.
\end{equation}%
The scale-factor $\Lambda _{QCD}^{5}$ enters on dimensional ground. In
virtue of the flavour content of the fields $\sigma _{B}[4q]=\frac{1}{2}%
[u,d][\overline{u},\overline{d}]$ and $f_{B}[4q]=\frac{1}{2\sqrt{2}}([u,s][%
\overline{u},\overline{s}]+[d,s][\overline{d},\overline{s}])$ and using the
parameter set of Eq. (\ref{solfulls}) together with $\Lambda _{QCD}\sim 0.25$
GeV we obtain:%
\begin{equation}
\left\vert \left\langle \frac{1}{2}[u,d][\overline{u},\overline{d}%
]\right\rangle \right\vert \sim \Lambda _{QCD}^{5}\beta _{3}\simeq 3.1\cdot
10^{-5}\text{ GeV}^{6}
\end{equation}%
\begin{equation}
\left\vert \left\langle \frac{1}{2}[u,s][\overline{u},\overline{s}%
]\right\rangle \right\vert =\left\vert \left\langle \frac{1}{2}[d,s][%
\overline{d},\overline{s}]\right\rangle \right\vert \sim \Lambda
_{QCD}^{5}\beta _{1}\simeq 1.9\cdot 10^{-5}\text{ GeV}^{6}
\end{equation}%
where the typical bare tetraquark masses $M_{\sigma _{B}[4q]}\sim 0.65$ GeV
and $M_{f_{B}[4q]}^{2}$ $\sim 1$ GeV have been employed \cite{tq}. The
precise value of the bare tetraquark masses is not relevant for our
estimation. It is interesting to notice that the magnitude of the
condensates is similar to \cite{fariborz}.

\subsubsection{Self-consistency problem}

The tetraquark nonet $\mathcal{S}^{[4q]}$ acquires non-zero $vev$'s $%
\left\langle \mathcal{S}_{ij}^{[4q]}\right\rangle =\beta _{i}\delta _{ij}$
(Eqs. (\ref{zeroval})-(\ref{tetvac})). Then, one has to shift the nonet as $%
\mathcal{S}_{ij}^{[4q]}\rightarrow \mathcal{S}_{ij}^{[4q]}+\beta _{i}\delta
_{ij}$ and substitute it back into the Lagrangian (\ref{lagd}). In
particular, when considering the mixing term of Eq. (\ref{lmixiso}) the
shift generates linear terms in the quarkonium scalar-isoscalar fields $N=%
\sqrt{\frac{1}{2}}(\overline{u}u+\overline{d}d)$ and $S=\overline{s}s$ :%
\begin{equation}
\mathcal{L}_{\text{mix-iso}}\rightarrow 2\sqrt{2}(c_{1}\alpha
_{s}+c_{2}\alpha _{u})N\beta _{1}+4\sqrt{2}(c_{1}\alpha _{u}+c_{2}\alpha
_{s})S\beta _{1}+2\sqrt{2}(c_{1}+c_{2})\alpha _{u}N\beta _{3}+...
\end{equation}%
Then, these linear terms modify the vacuum expectation values for the scalar
quarkonium nonet of Eq. (\ref{csb1}) as:%
\begin{equation}
\left\langle \mathcal{S}_{11}\right\rangle =\left\langle \mathcal{S}%
_{22}\right\rangle =\alpha _{u}\rightarrow \alpha _{u}+2\frac{(c_{1}\alpha
_{s}+c_{2}\alpha _{u})\beta _{1}+(c_{1}+c_{2})\alpha _{u}\beta _{3}}{%
M_{N}^{2}}  \label{alfaucorr}
\end{equation}%
\begin{equation}
\left\langle \mathcal{S}_{33}\right\rangle =\alpha _{s}\rightarrow \alpha
_{s}+\frac{4\sqrt{2}(c_{1}\alpha _{u}+c_{2}\alpha _{s})}{M_{S}^{2}}\beta _{1}
\label{alfascorr}
\end{equation}%
where $M_{N}$ and $M_{S}$ refer to the bare quarkonia masses; we employ the
typical values $M_{N}\sim 1.3$ GeV and $M_{S}\sim 1.6$ GeV \cite%
{giacosachiral}.

The modification of the vacuum expectation values of the scalar quarkonia
fields acknowledges the problem mentioned in Section 3.1: the minimum $%
\Sigma _{0}$ of the potential $V_{0}+V_{SB}$ is not anymore a minimum of our
entire potential $V_{0}+V_{SB}-\mathcal{L}_{\mathcal{S}^{[4q]}\Sigma \Sigma
}.$

We have to take it into account when evaluating the values of $\alpha _{u}$
and $\alpha _{s}$ from Eq. (\ref{alfaeq}). When using the modified
expressions (\ref{alfaucorr})-(\ref{alfascorr}) in the Eq. (\ref{csb1}),
here rewritten as

\begin{equation}
\left\langle \mathcal{S}_{11}\right\rangle +\left\langle \mathcal{S}%
_{22}\right\rangle =\sqrt{2}F_{\pi }\text{, }\left\langle \mathcal{S}%
_{11}\right\rangle +\left\langle \mathcal{S}_{33}\right\rangle =\sqrt{2}F_{K}%
\text{ ,}
\end{equation}%
the constants $\alpha _{u}$ and $\alpha _{s}$ change as follows (the
parameter set of Eq. (\ref{solfulls}) and the above listed scalar masses are
employed):%
\begin{equation}
\alpha _{u}=\frac{F_{\pi }}{\sqrt{2}}\rightarrow \frac{F_{\pi }}{\sqrt{2}}%
(1-0.11)\text{ ,}  \label{alfaunew}
\end{equation}

\begin{equation}
\alpha _{s}=-\frac{F_{\pi }}{\sqrt{2}}+\sqrt{2}F_{K}\rightarrow \left( -%
\frac{F_{\pi }}{\sqrt{2}}+\sqrt{2}F_{K}\right) (1-0.06)\text{ .}
\label{alfasnew}
\end{equation}

The corrections to the vacuum expectation values are $11\%$ and $6\%$
respectively, that is safely small. The results of the mixing evaluation in
Section 3.2 and in Appendix B are therefore confirmed. Indeed, the imprecise
knowledge of the experimental coupling constants of Eqs. (\ref{gqa0980})-(%
\ref{gqa01450}) generates a larger uncertainty than the neglect of the $vev$%
's corrections. The smallness of the latter originates from factors of the
kind $(\sqrt{2}\alpha _{u}/M_{scalar})^{2}\sim (F_{\pi }/M_{scalar})^{2}$ in
Eqs. (\ref{alfaucorr})-(\ref{alfascorr}), where $M_{scalar}\sim 1$ GeV
refers to the typical order of magnitude for the bare scalar tetraquark and
quarkonia fields.

One should then proceed iteratively, by shifting again the $vev$ of the
scalar-quarkonia fields according to Eqs. (\ref{alfaucorr})-(\ref{alfascorr}%
) and subsequently finding in the Lagrangian the new linear terms in the
scalar tetraquark fields (which arise from the scalar-tetraquark mixing
terms); in turn, this procedure allows to determine the next-to-leading
order correction to the tetraquark $vev$ of Eq. (\ref{tetvac}). For
instance, by using the new `alfa-values' of Eqs. (\ref{alfaunew})-(\ref%
{alfasnew}) and calculating the next-order correction to the $vev$ $\beta
_{3},$ a slight increase of $4\%$ is found, a small fraction which does not
change the results of the previous subsection about the tetraquark
condensates. This result is expected because the n-th correction involves
factors like $(F_{\pi }/M_{scalar})^{2n}$, thus decreasing very fast.
Naively, the minimum of the extended potential $V_{0}+V_{SB}-\mathcal{L}_{%
\mathcal{S}^{[4q]}\Sigma \Sigma }$ is close to the minimum $\Sigma _{0}$ of $%
V_{0}+V_{SB}.$ In the next subsection an explicit study of this issue by
means of a simple toy model is performed avoiding complicated algebraic
expressions. Notice that the iterative process is the unique way to proceed
because the exact form of the potential $V_{0}+V_{SB}$ is not specified.

The shift of the tetraquark nonet also induces contributions to the
pseudoscalar and scalar quarkonia masses (terms $\mathcal{L}_{\mathcal{S}%
^{[4q]}PP}$ and $\mathcal{L}_{\mathcal{S}^{[4q]}SS}$ in Eq. (\ref{decom})).
This fact has no influence in this work because we do not evaluate the bare
quarkonia and tetraquark masses.

\subsubsection{Toy potential}

Let us consider only the light quarks $u$ and $d$: as well known, chiral
symmetry invariance under $SU_{R}(2)\times SU_{L}(2)$ is fulfilled by
considering $\Sigma =N\tau ^{0}+i\pi ^{i}\tau ^{i},$ where $N\equiv 
\overline{n}n=\sqrt{1/2}(\overline{u}u+\overline{d}d)$ is the
isoscalar-quarkonium field, $\pi ^{i}$ the pseudoscalar pionic fields and $%
\tau ^{i}$ the 3 Pauli matrices ($\tau ^{0}$ is the $2\times 2$ identity
matrix).

In the $SU(2)$ limit only one scalar-diquark field survives: $\varphi =\sqrt{%
\frac{1}{2}}\varepsilon _{jk}q_{j}^{t}C\gamma ^{5}q_{k}.$ The scalar-diquark
matrix is given by $D=\varphi A$ where $A=i\tau ^{2}.$ Thus, we are left
with only one tetraquark field: $T=\sigma _{B}[4q]=\frac{1}{2}[u,d][%
\overline{u},\overline{d}].$

The Lagrangian $\mathcal{L}_{\mathcal{S}^{[4q]}\Sigma \Sigma }$ in Eq. (\ref%
{lagd}) reduces to the very simple form:%
\begin{equation}
\mathcal{L}_{\mathcal{S}^{[4q]}\Sigma \Sigma }\rightarrow gT\left( N^{2}+%
\overrightarrow{\pi }^{2}\right) ,\text{ }g=c_{1}+c_{2}.
\end{equation}%
(In the $SU(2)$ case the expressions for the dominant and subdominant terms
in large-$N_{c}$ expansion coincide). For illustrative purpose we use the
usual Mexican-hat potential $V_{0}=\frac{\lambda }{4}(N^{2}+\overrightarrow{%
\pi }^{2}-F^{2})^{2}$ and neglect $V_{SB}.$ Thus, the toy potential of the
reduced $SU(2)$-problem reads:%
\begin{equation}
V_{\text{toy}}=\frac{\lambda }{4}(N^{2}+\overrightarrow{\pi }^{2}-F^{2})^{2}+%
\frac{1}{2}M_{T}^{2}T^{2}-gT\left( N^{2}+\overrightarrow{\pi }^{2}\right)
\end{equation}%
where a mass-term for the tetraquark field has been included.

The minimum of $V_{0}$ is at $\{N_{0}=F,\overrightarrow{\pi }=%
\overrightarrow{0}\}.$ By expanding around this point, i.e. shifting $%
N\rightarrow F+N,$ the quantity $\mathcal{L}_{\mathcal{S}^{[4q]}\Sigma
\Sigma }$ in the SU(2) limit generates analogous terms to those discussed
throughout this section: 
\begin{equation}
\mathcal{L}_{\mathcal{S}^{[4q]}\Sigma \Sigma }\rightarrow gT\left( N^{2}+%
\overrightarrow{\pi }^{2}\right) =gT(N^{2}+\overrightarrow{\pi }%
^{2}+2FN+F^{2})  \label{lagdsu(2)}
\end{equation}%
In fact, we recognize the tetraquark-mesons decay terms, the mixing term
(whose strength amounts to $2gF$) and the linear term in the field \ $T$.

But the minimum of $V_{0}$ is not the minimum of $V_{\text{toy}}.$ The
corresponding minimum point of $V_{\text{toy}}$, denoted as $P_{\min
}=\{N_{0},T_{0},\overrightarrow{\pi }=\overrightarrow{0}\},$ can be
analytically calculated:%
\begin{eqnarray}
N_{0} &=&\frac{F}{\sqrt{1-\frac{2g^{2}}{\lambda M_{T}^{2}}}}=F\left( 1+\frac{%
g^{2}}{\lambda M_{T}^{2}}+...\right) ,\text{ } \\
T_{0} &=&\frac{g}{M_{T}^{2}}N_{0}^{2}=\frac{g}{M_{T}^{2}}\frac{F^{2}}{1-%
\frac{2g^{2}}{\lambda M_{T}^{2}}}=\frac{gF^{2}}{M_{T}^{2}}\left( 1+\frac{%
2g^{2}}{\lambda M_{T}^{2}}+...\right) .
\end{eqnarray}%
When $g=0$ we reobtain the minimum at $N_{0}=F$ and $T_{0}=0.$ If the term $%
g^{2}/\lambda M_{T}^{2}$ entering in the expansions is small, one simply has
small corrections to the value $N_{0}=F$ . The toy-model clarifies what kind
of correction terms one evaluates in the iterative process sketched in the
previous subsection leading to Eqs. (\ref{alfaunew})-(\ref{alfasnew}).

The bare mass of the field $N$ and the mixing strength can be also exactly
evaluated by expanding around the minimum $P_{\min }$:%
\begin{equation}
M_{N}^{2}=\left( \frac{\partial ^{2}V_{\text{toy}}}{\partial N^{2}}\right)
_{P=P_{\min }}=\frac{2\lambda F^{2}}{1-\frac{2g^{2}}{\lambda M_{T}^{2}}},%
\text{ }\left( \frac{\partial ^{2}V_{\text{toy}}}{\partial N\partial T}%
\right) _{P=P_{\min }}=-2gN_{0}.
\end{equation}

Let us estimate the corrections. The parameter $\lambda $ at first order is: 
$\lambda \simeq M_{N}^{2}/2F^{2}.$ This value is accurate if $\frac{2g^{2}}{%
\lambda M_{T}^{2}}\simeq \frac{4g^{2}F^{2}}{M_{T}^{2}M_{N}^{2}}<<1.$ The
condition is satisfied in our case. In fact, using the typical values $g\sim
1.5$ GeV (as in Eq. (\ref{solfulls})) $F\sim F_{\pi },$ $M_{T}\sim 0.65$ GeV
and $M_{N}\sim 1.3$ GeV one has $\frac{4g^{2}F_{\pi }^{2}}{M_{T}^{2}M_{N}^{2}%
}\simeq 0.1$; we also see the appearance of factor like $(F_{\pi
}/M_{scalar})^{2}$ in the Taylor expansion, as already discussed in Section
3.2.

\section{Summary and conclusions}

This work aimed to study the implications of Jaffe's tetraquark states as a
necessary component to correctly interpret the scalar low-energy QCD sector.
We summarize the relevant points.

a) The scalar and the pseudoscalar quarkonia nonets are introduced in the
usual fashion. We did not specify the potential for these fields, but we
solely assumed chiral symmetry breaking to occur, thus non-vanishing $vev$'s
for the isoscalar quarkonia fields, in turn related to the pion and kaon
decay constants $F_{\pi }$ and $F_{K},$ are generated. The bare scalar
quarkonia masses are set above 1 GeV (in accord with the Lattice study of
Ref. \cite{prelovsek,degrand,mathur}), where the other p-wave nonets of
axial-vector and tensor mesons lie.

b) The scalar diquark in the flavor $\left( \overline{3}_{F}\right) $and
color $\left( \overline{3}_{C}\right) $ antitriplet configurations is a
compact and stable object, thus a good candidate for the basic building
block of the light scalar mesons, which naturally emerge as a tetraquark
scalar nonet. This assignment is in agreement with the mass-degeneracy of $%
a_{0}(980)$ and $f_{0}(980)$, their large $\overline{K}K$ decay strengths
and their non-quarkonia behavior for large-$N_{c}$ analysis. These facts,
together with point (a), support the bare level ordering $M_{4q}<M_{%
\overline{q}q}.$

c) In a chiral framework the $\left( \overline{3}_{F},\overline{3}%
_{C}\right) $ pseudoscalar diquark is introduced as the chiral partner of
the scalar diquark. Chiral symmetry breaking driven by instantons predicts a
strong attraction in the scalar channel and a repulsion in the pseudoscalar
one. This fact makes the pseudoscalar diquark heavier and loosely bound,
thus we do not consider it as a relevant constituent for the light meson
spectroscopy. For instance, an extra non-quarkonia nonet built out two
pseudoscalar diquark is not seen in the spectrum below 2 GeV.

d) A tetraquark-quarkonia interaction Lagrangian invariant under $%
SU_{R}(3)\times SU_{L}(3)\times U_{A}(1)$ is written down at the leading and
subleading order in the large-N$_{c}$ expansion. Both scalar and
pseudoscalar diquark constituents enter in its expression. Then, in virtue
of point (c) only the scalar diquark and the corresponding tetraquark nonet
are taken into account.

e) The $\chi SB$ of point (a) generates a mixing term among the scalar
quarkonia and tetraquark nonet. The corresponding mixing strengths are a
linear combination of $F_{\pi }$ and $F_{K}$ and the tetraquark decay
strengths $c_{1}$ and $c_{2},$ which parametrize the processes of Figs. 1.a
and 1.b. The mixing is then evaluated in the isovector channel: $a_{0}(980)$
is mainly a Jaffe's tetraquark state, with a small quarkonium amount $%
(\lesssim 10\%),$ and $a_{0}(1450)$ has a dominant quarkonium content. The
results are similar in the kaonic sector and are stable under changes of the
employed parameters, as long as the bare level ordering $M_{4q}<M_{\overline{%
q}q}$ holds.

f) The $\chi SB$ at a quarkonia level induces also linear terms in the
isoscalar tetraquark fields, thus non-vanishing $vev$'s for the latter
emerge. They are also related to the magnitude of corresponding four-quark
condensate(s), whose values have been estimated about $2$-$3\cdot 10^{-5}$
GeV$^{6}$. As a last step a self-consistency check about the minimum of
scalar-isoscalar fields has been done and a simple toy-model for the reduced 
$SU(2)$ problem discussed.

We found a substantial separation of the tetraquark states (below 1 GeV) and
quarkonia states (between 1-2 GeV, where the scalar glueball intrudes in the
isoscalar sector). The confirmation of the falsification of this scenario is
an important issue of low-energy hadronic QCD. Furthermore, decays of heavy
states in the charmonia region involve the scalar mesons below 2 GeV. Thus,
the correct interpretation of the latter is a crucial step for the analysis
of the decays of charmonia and heavy-glueball states, which according to
Lattice QCD are believed to show up in the mass region between 3-5 GeV \cite%
{morningstar}, in turn related to the planed experimental search of PANDA at
FAIR \cite{peters}.

As an interesting development, the analysis of electromagnetic decay of (and
into) vector meson such as $V\rightarrow S^{[4q]}\gamma $ \cite{morgan} and $%
S^{[4q]}\rightarrow V\gamma $ \cite{kalashnikova} within a phenomenological
composite Lagrangian can constitute a useful step in disentangling the
nature of the light scalar states below 1 GeV and is planned as a future
work. Along the same line, possible interactions involving the
experimentally well-known tensor mesons within a composite approach as in 
\cite{tensor} can also be performed.

\section*{Acknowledgments}

The author thanks D. Rischke for stimulating and useful discussions and J.
R. Pelaez for helpful remarks to the first version of this manuscript.

\appendix

\section{ Nonets and their transformation}

Out of the introduced diquarks we can (formally) identify 4 nonets, with
definite properties under chiral transformations. We first consider the
matrix of tetraquark states $T$ (analogous to $\widehat{\Phi }$ in \cite%
{napsuciale} and $M^{\prime }$ in \cite{fariborz}):%
\begin{equation}
T_{ij}=\sqrt{2}\varphi _{i}^{L\dagger }\varphi _{j}^{R};\text{ }%
T=T^{S}+iT^{P},\text{ }
\end{equation}%
which constitutes of a scalar and a pseudoscalar nonet of tetraquark states
given by the Hermitian matrices: 
\begin{equation}
T_{ij}^{S}=\frac{1}{\sqrt{2}}\left( \widetilde{\varphi }_{i}^{\dagger }%
\widetilde{\varphi }_{j}-\varphi _{i}^{\dagger }\varphi _{j}\right) ,\text{ }%
T_{ij}^{P}=\frac{i}{\sqrt{2}}\left( \varphi _{i}^{\dagger }\widetilde{%
\varphi }_{j}-\widetilde{\varphi }_{i}^{\dagger }\varphi _{j}\right) \text{ .%
}
\end{equation}%
The matrices $T,$ $T^{S}$ and $T^{P}$ transform as $\Sigma ,$ $S$ and $P$ in
Table 1, except for the $U_{A}(1)$ transformation, which now reads $%
T\rightarrow e^{-4i\nu }T$. In particular, for chiral transformations: $%
T\rightarrow LTR^{\dagger }$. Notice that the scalar nonet $\mathcal{S}%
^{[4q]}$ of Eqs. (\ref{s4q}) and (\ref{s4q2}) is now a part of $T^{S}$. In
the chiral context the scalar nonet $T^{S}$ is an admixture of both
diquarks. In \cite{black,napsuciale} the tetraquark-quarkonia mixing occurs
via the chirally invariant (but not $U_{A}(1)$ invariant) term 
\begin{equation}
\mathcal{L}_{\text{mix}}=e\cdot Tr[\Sigma T^{\dagger }+T\Sigma ^{\dagger }]
\end{equation}%
where $e$ is a free parameter.

Other two tetraquark meson nonets can be formed: 
\begin{equation}
T_{ij}^{R}=\sqrt{2}\varphi _{i}^{R\dagger }\varphi _{j}^{R},\text{ }%
T_{ij}^{L}=\sqrt{2}\varphi _{i}^{L\dagger }\varphi _{j}^{L}  \label{rl}
\end{equation}%
which under chiral transformations transform as $T^{R}\rightarrow
RT^{R}R^{\dagger }$ and $T^{L}\rightarrow LT^{R}L^{\dagger },$ i.e. such as
quark's left and right currents, connected to vector and axial-vector
mesons. In the present context we still deal with scalar and pseudoscalar
tetraquark states, which we denote as $\Pi ^{S}$ and $\Pi ^{P}$:%
\begin{eqnarray}
\Pi ^{S} &=&\frac{1}{2}(T^{R}+T^{L});\text{ }\Pi _{ij}^{S}=\frac{1}{\sqrt{2}}%
\left( \widetilde{\varphi }_{i}^{\dagger }\widetilde{\varphi }_{j}+\varphi
_{i}^{\dagger }\varphi _{j}\right) \\
\Pi ^{P} &=&\frac{1}{2}(T^{R}-T^{L});\text{ }\Pi _{ij}^{P}=\frac{1}{\sqrt{2}}%
\left( \varphi _{i}^{\dagger }\widetilde{\varphi }_{j}+\widetilde{\varphi }%
_{i}^{\dagger }\varphi _{j}\right)
\end{eqnarray}%
The scalar and pseudoscalar nonets $\Pi ^{S}$ and $\Pi ^{P}$ transform as
vector and axial-vector under chiral transformation. Out of a quark and an
antiquark such scalar and pseudoscalar objects do not exist because they
vanish identically (direct product $P_{R}\cdot P_{L}=0$ in the expression
for the currents). They are however possible for tetraquark states: four
nonets can be then formed.

After chiral symmetry breaking at a diquark level occurs, there is no reason
that the physical nonets are those listed in the present Appendix. The
scalar nonets $T^{S}$ and $\Pi ^{S}$ can mix and split. This fact resembles
the flavour wave functions of the vector mesons $\omega $ and $\phi $, where
the quark mass splitting generates a separation of $(u,d)$ and $s$ quark
dynamics. We assumed that the splitting is large enough to generate two
separated nonets of scalar and pseudoscalar diquark constituents:%
\begin{equation}
T^{S},\text{ }\Pi ^{S}\rightarrow \mathcal{S}^{[4q]}=\varphi _{i}^{\dagger
}\varphi _{j},\text{ }\widetilde{\varphi }_{i}^{\dagger }\widetilde{\varphi }%
_{j}
\end{equation}%
\qquad The scalar nonet $\widetilde{\varphi }_{i}^{\dagger }\widetilde{%
\varphi }_{j}$ does not show up in the spectrum below 2 GeV. It could be
heavier, too broad or simply not realized in nature. Here we simply
concentrated on $\mathcal{S}^{[4q]}.$ A more quantitative analysis of the
splitting of scalar and pseudoscalar diquarks would represent an interesting
subject on its own.

\section{Results for parameter variation}

We evaluate the quarkonium amount in the resonance $a_{0}(980),$ represented
by the quantity $\sin ^{2}\theta ,$ for different choices of the parameters.
We consider all three values for $g_{_{a_{0}(980)\rightarrow \eta \pi
}}^{2}=5$, $7.5$, $10$ GeV$^{2}$ listed in Eq. (\ref{gqa0980}). We first
employ $g_{_{a_{0}(1450)\rightarrow \eta \pi }}^{2}=10.4$ GeV$^{2}$ and we
consider different values for the ratio $c_{2}/c_{1}.$ (The value $%
c_{2}/c_{1}=1$ would imply large-$N_{c}$ violation. Here it is used to show
the stability of the results under changes of this ratio).

\begin{center}
\textbf{Table 4: } $\sin ^{2}\theta $ when varying $g_{_{a_{0}(980)%
\rightarrow \eta \pi }}^{2}$ and $\frac{c_{2}}{c_{1}}.$

\ \ \ \ \ \ \ \ \ \ \ \ \ \ \ \ \ \ \ \ \ \ \ \ ($g_{_{a_{0}(1450)%
\rightarrow \eta \pi }}^{2}=10.4$ GeV$^{2}$)\ \ \ \ \ \ \ \ \ \ \ \ \ \ \ 

$%
\begin{tabular}{|c|c|c|c|}
\hline
& ($\frac{c_{2}}{c_{1}}=\frac{1}{3}$) & ($\frac{c_{2}}{c_{1}}=0.62$) & ($%
\frac{c_{2}}{c_{1}}=1$) \\ \hline
$g_{_{a_{0}(980)\rightarrow \eta \pi }}^{2}$ (GeV$^{2}$) & $\sin ^{2}\theta $
& $\sin ^{2}\theta $ & $\sin ^{2}\theta $ \\ \hline
5 & 3.23\% & 3.22\% & 3.22\% \\ \hline
7.5 & 4.82\% & 4.81\% & 4.80\% \\ \hline
10 & 6.40\% & 6.39\% & 6.38\% \\ \hline
\end{tabular}%
$
\end{center}

Notice that the dependence on the ratio $c_{2}/c_{1}$ is extremely weak.

As stressed in\ Section 3.2.2 the value for $g_{_{a_{0}(1450)\rightarrow
\eta \pi }}^{2}$ can be smaller than $10.4$ GeV$^{2}.$ We evaluate $\sin
^{2}\theta $ for $g_{_{a_{0}(1450)\rightarrow \eta \pi }}^{2}=5.2$ GeV$^{2}.$
This value corresponds to a width four times smaller: $\Gamma
_{a_{0}(1450)\rightarrow \eta \pi }\simeq 119/4=29.75$ MeV, probably to
small. It can be regarded as a lower limit.

\begin{center}
\textbf{Table 5: } $\sin ^{2}\theta $ when varying $g_{_{a_{0}(980)%
\rightarrow \eta \pi }}^{2}$ and $\frac{c_{2}}{c_{1}}.$

\ \ \ \ \ \ \ \ \ \ \ ($g_{_{a_{0}(1450)\rightarrow \eta \pi }}^{2}=5.2$ GeV$%
^{2}$)

$%
\begin{tabular}{|c|c|c|c|}
\hline
& ($\frac{c_{2}}{c_{1}}=\frac{1}{3}$) & ($\frac{c_{2}}{c_{1}}=0.62$) & ($%
\frac{c_{2}}{c_{1}}=1$) \\ \hline
$g_{_{a_{0}(980)\rightarrow \eta \pi }}^{2}$ (GeV$^{2}$) & $\sin ^{2}\theta $
& $\sin ^{2}\theta $ & $\sin ^{2}\theta $ \\ \hline
5 & 3.79\% & 3.78\% & 3.77\% \\ \hline
7.5 & 5.66\% & 5.65\% & 5.63\% \\ \hline
10 & 7.51\% & 7.50\% & 7.48\% \\ \hline
\end{tabular}%
$
\end{center}

The previous results are confirmed.

As a last step we include a possible momentum dependence for the quarkonium
coupling constant: we use the dominant term of Ref. \cite{giacosachiral}
obtained in the framework of a chiral Lagrangian: $g_{_{a_{0}[\overline{q}%
q]\rightarrow \eta \pi }}=\gamma (p^{2}-M_{\pi }^{2}-M_{\eta }^{2})$ where $%
\gamma $ is a constant involving the pseudoscalar angle $\theta _{P}$ and a
nonet decay strength. Strictly speaking one should consistently also include
a running coupling constant for the four-quark amplitude $%
g_{_{a_{0}[4q]\rightarrow \eta \pi }}$. This operation goes beyond the goal
of this work. The present aim is to show the stability of the results even
in presence of an explicit momentum-dependence of the amplitude $g_{_{a_{0}[%
\overline{q}q]\rightarrow \eta \pi }}$.

\begin{center}
\textbf{Table 6: }$\sin ^{2}\theta $ when varying $g_{_{a_{0}(980)%
\rightarrow \eta \pi }}^{2}$ and $\frac{c_{2}}{c_{1}}.$

\ \ \ \ \ \ \ \ \ \ \ (running $g_{_{a_{0}[\overline{q}q]\rightarrow \eta
\pi }}$)

$%
\begin{tabular}{|c|c|c|c|}
\hline
& ($\frac{c_{2}}{c_{1}}=\frac{1}{3}$) & ($\frac{c_{2}}{c_{1}}=0.62$) & ($%
\frac{c_{2}}{c_{1}}=1$) \\ \hline
$g_{_{a_{0}(980)\rightarrow \eta \pi }}^{2}$ (GeV$^{2}$) & $\sin ^{2}\theta $
& $\sin ^{2}\theta $ & $\sin ^{2}\theta $ \\ \hline
5 & 4.97\% & 4.96\% & 4.94\% \\ \hline
7.5 & 7.68\% & 7.66\% & 7.64\% \\ \hline
10 & 10.56\% & 10.53\% & 10.51\% \\ \hline
\end{tabular}%
$
\end{center}

The results point to a slightly larger quarkonium content, which is however
still smaller than $11\%.$ Furthermore, this value is realized for $%
g_{_{a_{0}(980)\rightarrow \eta \pi }}^{2}=10$ GeV$^{2}$, which, as
discussed in Section 3.2.2, can be considered as an upper limit.

\newpage

\begin{figure}[tbp]
\begin{center}
\leavevmode
\leavevmode
\vspace{4.3cm} \includegraphics{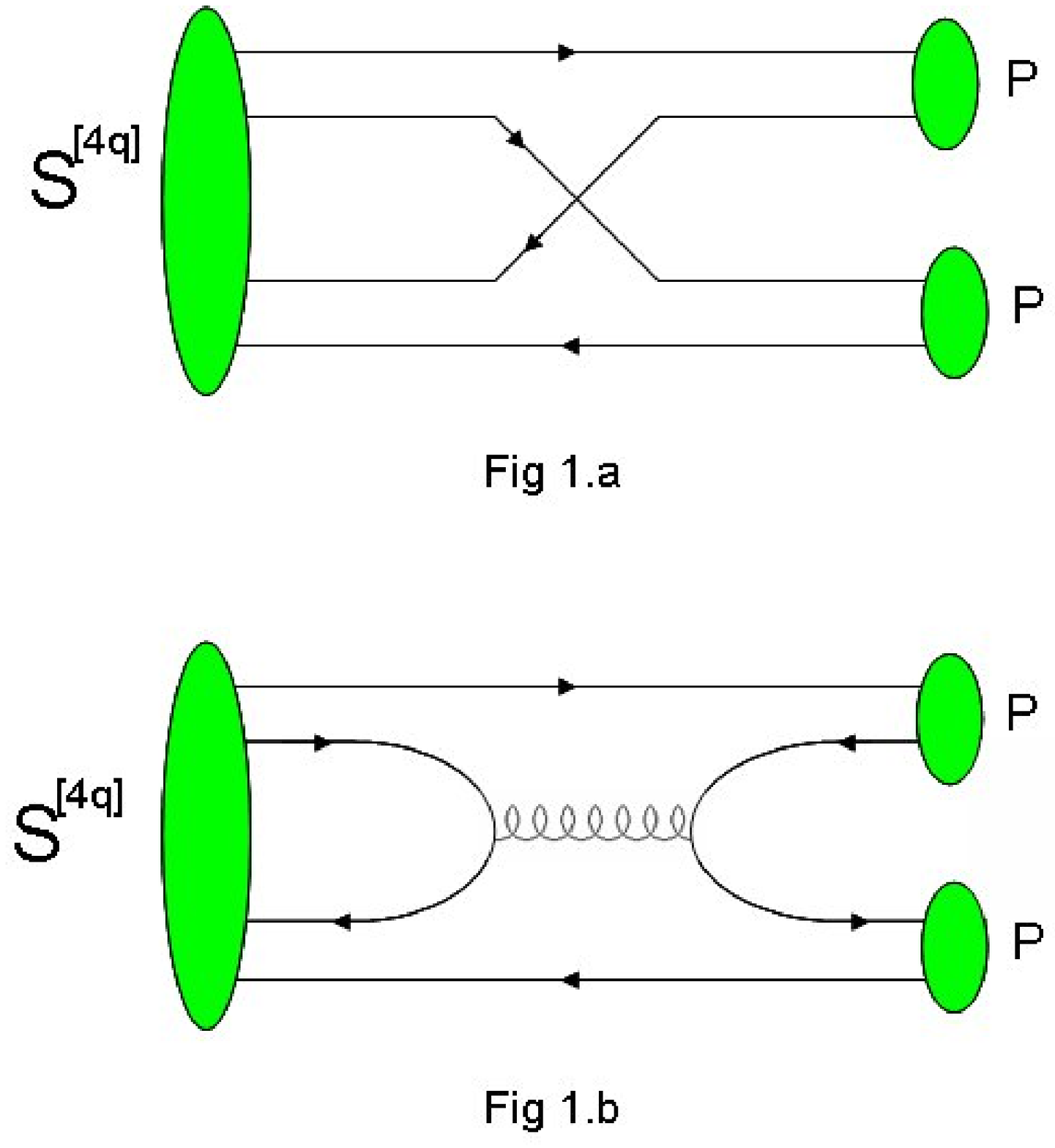}
\end{center}
\par
\vspace{5.4cm}
\caption{{Dominant (1.a) and subdominant (1.b) contributions to the
transition amplitudes of a scalar tetraquark state into two pseudoscalar
mesons. They correspond to the Lagrangian in Eqs. (\protect\ref{ltpp})-(%
\protect\ref{lttp2}).}}
\label{Fig1}
\end{figure}

\begin{figure}[tbp]
\begin{center}
\leavevmode
\leavevmode
\vspace{4.3cm} \includegraphics{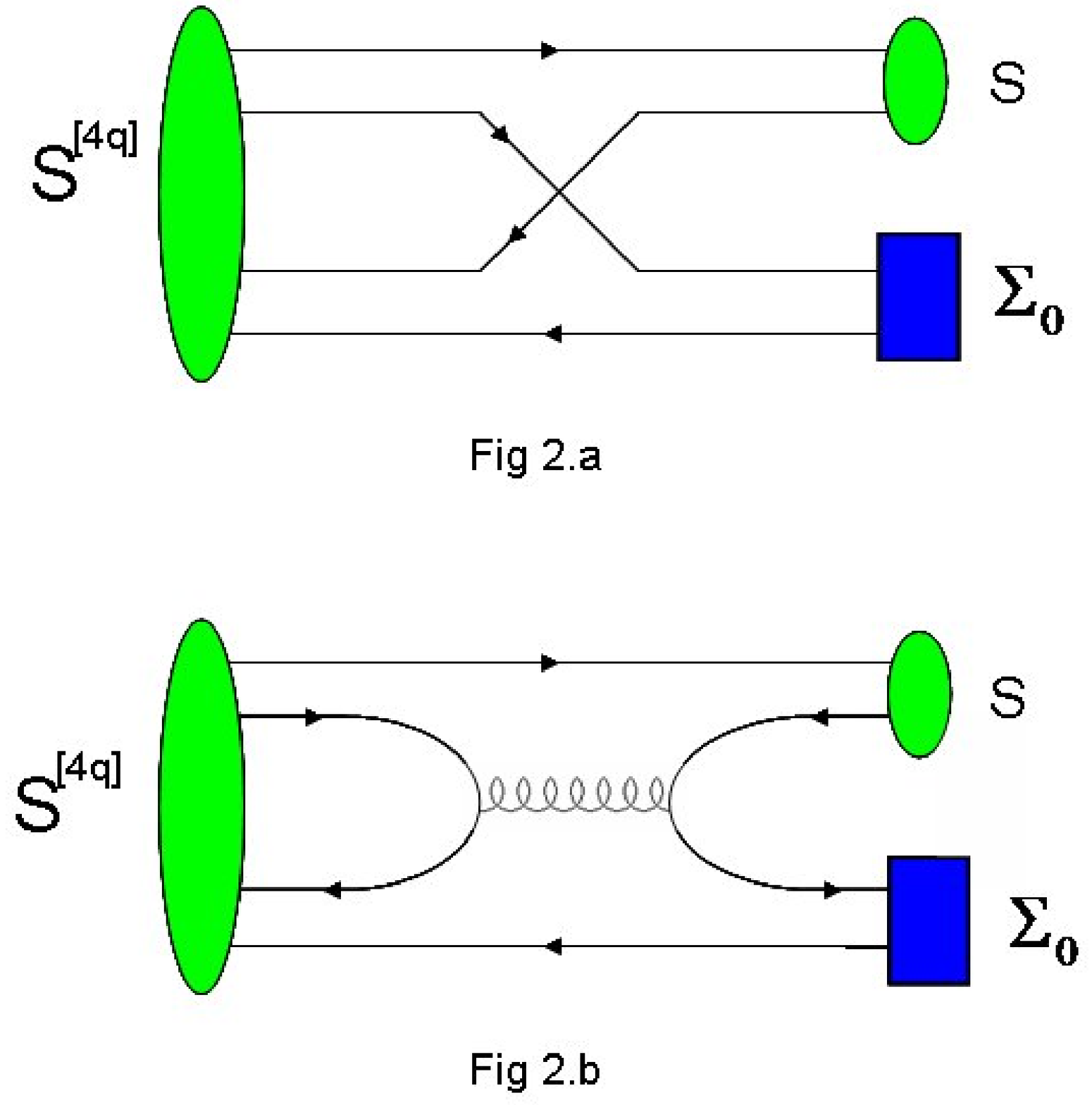}
\end{center}
\par
\vspace{5.4cm}
\caption{{Dominant (2.a) and subdominant (2.b) contributions to the
tetraquark-quarkonia mixing, corresponding to Eq. (\protect\ref{lmix}). At
one vertex the $vev$'s for the scalar-isoscalar quarkonia fields, encoded in
the matrix $\Sigma _{0}$ of Eq. (\protect\ref{sigma0}), appears. As a
consequence a tetraquark-quarkonia mixing is generated, whose strength is
related to $\Sigma _{0}$, i.e. to the pion and kaon decay constants.}}
\label{Fig2}
\end{figure}

\begin{figure}[tbp]
\begin{center}
\leavevmode
\leavevmode
\vspace{4.3cm} \includegraphics{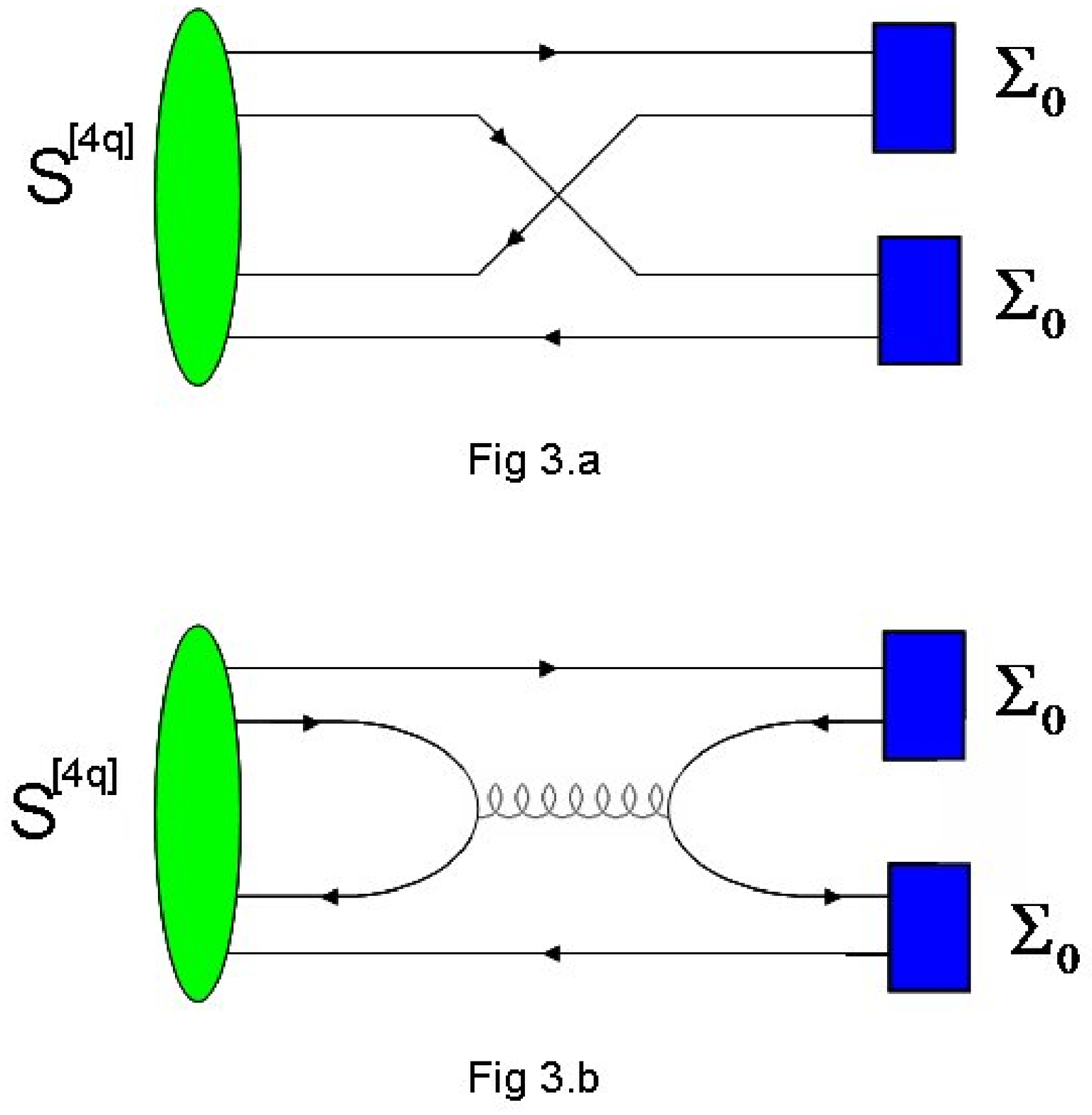}
\end{center}
\par
\vspace{5.4cm}
\caption{{Dominant (3.a) and subdominant (3.b) contributions to the linear
terms in the scalar-isoscalar tetraquark fields, which generate a non-zero
vacuum expectaion value for the latter, see Eq. (\protect\ref{lconds}). At
both vertices the the $vev$'s of the scalar-isoscalar quarkonia fields,
encoded in the matrix $\Sigma _{0}$ of Eq. (\protect\ref{sigma0}), appear.}}
\label{Fig3}
\end{figure}

\end{document}